\documentclass[a4paper,11pt]{article}
\pdfoutput=1 

\usepackage{jheparxiv}

\usepackage[T1]{fontenc} 

\usepackage{tensor}
\usepackage{amsmath}
\usepackage{amssymb}
\usepackage{mathrsfs}
\usepackage{mathtools}
\usepackage{bbm}
\usepackage{graphicx}
\usepackage{stmaryrd}
\usepackage{twistor}
\usepackage{empheq}
\usepackage{cancel}

\usepackage{soul}

\renewcommand{\sc}{\mathsf{c}}

\newtheorem{theorem}{Theorem}

\newcommand{\qed}{\hfill \ensuremath{\Box}}

\def\d{{\rm d}}
\def\i{{\rm i}}

\title{The dual twistor theory of self-dual black holes}

\author[a]{Tim Adamo,}
\author[a,b]{Bernardo Araneda}
\author[a]{\& Sean Seet}

\affiliation[a]{School of Mathematics and Maxwell Institute for Mathematical Sciences,\\
University of Edinburgh, EH9 3FD, United Kingdom\vspace{0.1cm}}
\affiliation[b]{Max-Planck-Institut f\"ur Gravitationsphysik (Albert-Einstein-Institut), \\
Am M\"uhlenberg 1, D-14476 Potsdam, Germany\vspace{0.1cm}}
\emailAdd{t.adamo@ed.ac.uk}
\emailAdd{baraneda@ed.ac.uk}
\emailAdd{sseet@ed.ac.uk}

\abstract{The Taub-NUT and Eguchi-Hanson gravitational instantons, along with the self-dual Pleba\'nski-Demia\'nski metric, form a set of Euclidean metrics which can naturally be called `self-dual black holes', as they arise from self-dual slices of the most general vacuum, asymptotically flat black hole metric. These self-dual black holes are of interest for many reasons, and can famously be described through the non-linear graviton construction of twistor theory. However, the implicit nature of this twistor description obscures some features of the underlying geometry, particularly for the most general self-dual black holes. In this paper, we give a new construction of all asymptotically flat self-dual black holes based on holomorphic quadrics in flat dual twistor space, rather than the usual twistor space associated with self-duality. Remarkably, the geometry of the self-dual black holes -- including their hyperk\"ahler structure, as well as Kerr-Schild and Gibbons-Hawking forms -- is directly encoded in the corresponding quadric. As a consequence, we obtain a previously unknown single Kerr-Schild form of the self-dual Pleba\'nski-Demia\'nski metric.}  

\begin{document}
\maketitle
\flushbottom

\section{Introduction}

Astrophysical black holes are remarkably simple objects, characterised by a small collection of charges (mass, spin and charge) thanks to the no-hair theorem~\cite{Chrusciel:2012jk}. Furthermore, they share in common the important feature of being algebraically special of type D in the Petrov-Pirani-Penrose classification: this means that their Weyl curvature tensor has two repeated principal null directions (cf., \cite{Penrose:1986ca}). The type D property is the underlying reason why it is possible to determine these metrics as exact solutions of the Einstein equations, and is intimately tied to the remarkable simplicity of the Teukolsky system~\cite{Teukolsky:1973ha} describing linearised gravitational perturbations to black hole spacetimes.

There are many imperatives to understand gravitational scattering in black hole metrics, ranging from gravitational wave astronomy to the holographic principle, which require fine-grained control over gravitational perturbations to black holes and their dynamics. However, despite the simplicity of black hole metrics and the Teukolsky equations, this remains a very difficult problem to approach analytically. For instance, to consider wave-wave scattering in the presence of a black hole requires analytic solutions to the Teukolsky equations as well as a mechanism to implement dynamics between them. The standard framework for performing these computations through the background field theory of the Einstein-Hilbert action is woefully complicated, and most modern efforts in this direction treat the black hole background perturbatively in some scheme (cf., \cite{Kosower:2022yvp,Buonanno:2022pgc}).

This motivates looking for alternative approaches to describing black holes and their perturbation theory which renders analytic calculations more tractable. One such approach is to consider toy models of astrophysical black holes which enable high-precision gravitational scattering calculations while still retaining the rich non-linearity of perturbation theory in a curved metric to such an extent that information about the full-blown (non-toy) scattering problem could eventually be extracted. A recent effort in this direction considers gravitational scattering on \emph{self-dual}, or \emph{hyperk\"ahler}, metrics: solutions of the vacuum Einstein equations with a self-dual Weyl curvature tensor. Such solutions cannot be Lorentzian-real, so one typically works in Euclidean or split/hyperbolic signature to obtain real manifolds. 

Simplification arises from the fact that the self-dual Einstein equations are classically integrable~\cite{Penrose:1976js,Boyer:1985aj}. Self-dual metrics and their perturbations have a description in terms of twistor theory~\cite{Penrose:1976js,Atiyah:1978wi,Hitchin:1980hp,Gindikin:1986}, and dynamics can be implemented (at least in some scenarios) via a two-dimensional chiral sigma model with twistor space as its target~\cite{Adamo:2021bej}. This has made it possible to obtain explicit formulae for graviton scattering amplitudes in a variety of self-dual backgrounds which are entirely beyond the reach of traditional, background-field methods (even in the self-dual setting)~\cite{Adamo:2020syc,Adamo:2022mev,Bittleston:2023bzp,Adamo:2023zeh,Adamo:2023fbj,Guevara:2023wlr,Bogna:2024gnt,Guevara:2024edh,Adamo:2025fqt,Guevara:2025psg}. However, at present it remains unclear how to `bootstrap' these remarkable results on self-dual backgrounds to learn something about scattering on Lorentzian-real, astrophysical backgrounds like the Kerr black hole.

\medskip

On the other hand, it was recently shown that all astrophysical, Lorentzian-real black hole metrics can be obtained by considering deformations of holomorphic, quadric hypersurfaces in the \emph{dual} twistor space of Minkowski spacetime -- the projective dual of the usual twistor space~\cite{Araneda:2022xii}. This construction exploits the fact that these metrics are all conformally K\"ahler~\cite{Flaherty:1974,Flaherty:1976,Aksteiner:2022bwr}, is intimately tied to the type D property and shares many of the geometric underpinnings with the existence of the Teukolsky system for perturbations of these spacetimes~\cite{Araneda:2018ezs,Araneda:2019uwy,Araneda:2021wcd}. This suggests an alternative approach to black hole and their perturbation theory which harnesses the hidden complex geometry of black holes. However, this has yet to be realized.

This motivates looking for a place where these two novel approaches to describing black holes and their perturbation theory -- namely, self-duality versus dual twistor quadrics -- overlap. The former exploits self-duality to make remarkable progress in perturbation theory, but without any obvious connection to the Lorentzian-real setting or a particular preference for algebraically type D backgrounds. The latter manifests the conformal K\"ahler geometry of type D metrics and holds for Lorentzian signature, but the way in which perturbation theory is implemented in the dual twistor quadric is obscure. Consequently, it seems clear that \emph{self-dual black holes} are a natural place to see where the best aspects of each approach come to the fore.

\medskip

In this paper, we develop a novel description of all self-dual black holes as holomorphic quadrics (i.e., holomorphic algebraic varieties of degree two) in an open subset of the complex projective space $\P^3$. Our precise notion of a self-dual black hole is any self-dual, Ricci-flat (i.e., hyperk\"ahler) Riemannian metric which arises as a Euclidean-real slice of the complexification of the most general type D black hole metric in four-dimensions: the Pleba\'nski-Demia\'nski metric~\cite{Plebanski:1976gy}. The self-dual Taub-NUT metric~\cite{Hawking:1976jb} -- which has been used as a self-dual analogy of a Schwarzschild or Kerr black hole previously in the literature~\cite{Crawley:2023brz,Adamo:2023fbj,Guevara:2023wlr} -- is one example, but the Eguchi-Hanson~\cite{Eguchi:1978xp,Eguchi:1978gw} and most general self-dual Pleba\'nski-Demia\'nski metrics are also included.

Since these are hyperk\"ahler metrics, they admit a twistor description through the non-linear graviton construction which is tied to their self-duality. In the cases of self-dual Taub-NUT and Eguchi-Hanson, these were known long ago~\cite{Hitchin:1979rts,Tod:1979tt,Sparling:1981nk,Tod:1982mmp}, although the case of SDPD seems less well-studied. Our goal is to develop an alternative description of self-dual black hole metrics which foregrounds the type D property and does not involve a deformation (e.g., of the complex structure of a twistor space) or require the solution of an associated linear problem (e.g., solving for holomorphic twistor curves).

A metric being algebraically type D endows it with a conformal K\"ahler structure~\cite{Flaherty:1974}, implying that the metric admits a (non-constant) valence-two Killing spinor. It has long been known that the existence of such Killing spinors is encoded by certain holomorphic surfaces in twistor space: in the flat case, this is the statement that a valence-$k$ Killing spinor corresponds to the zero set of a degree $k$ homogeneous, holomorphic polynomial in twistor space (i.e., to a holomorphic algebraic variety of degree $k$)~\cite{Penrose:1967wn}. Similar results hold for curved self-dual 4-manifolds~\cite{Tod:1979tt,Pontecorvo:1992}. Consequently, one might expect that type D self-dual black hole metrics will have special quadrics (i.e., quadratic holomorphic hypersurfaces) in their associated twistor spaces.

\medskip

Instead, we find a construction of all self-dual black hole metrics which utilizes only quadrics in the \emph{dual} twistor space of \emph{flat} space. This is remarkable for several reasons: firstly, that it is the dual twistor space -- typically associated with \emph{anti}-self-dual degrees of freedom -- rather than twistor space which underpins the construction, and secondly that one requires only the dual twistor space of flat space to obtain curved, self-dual black hole metrics. Furthermore, the construction is substantially more explicit than the standard non-linear graviton construction: all aspects of the metric and its hyperk\"ahler geometry are read off directly from the dual twistor quadric. 

Our construction can be stated as an algorithm, which proceeds as follows:
\begin{enumerate}
 \item Let $Q$ be a generic holomorphic quadric in the dual twistor space $\PT^*$ of complexified Minkowski space, compatible with Euclidean reality conditions. 
 \item Via the Penrose transform~\cite{Penrose:1969ae}, $Q$ gives rise to a null, self-dual Maxwell field $\varphi_{\dot\alpha\dot\beta}(x)$ on flat space.
 \item A remarkable theorem of Tod~\cite{Tod:1982mmp} constructs a Kerr-Schild hyperk\"ahler metric from this null, self-dual Maxwell field. 
\end{enumerate}
In this way, every dual twistor quadric gives rise to a hyperk\"ahler metric in Kerr-Schild form. Furthermore, various other geometric structures -- including a valence-2 Killing spinor, various Killing vectors and tensors, and a basis of anti-self-dual two forms -- are built directly from the ingredients in the dual twistor quadric. These can be used to show that every metric arising via this algorithm also admits a Gibbons-Hawking form.

It is then possible to classify all such dual twistor quadrics, and study each case in turn. We show that there are precisely three classes of dual twistor quadric, and each one leads to a type D hyperk\"ahler metric of a particular form. Putting everything together gives our main result:
\begin{theorem}\label{thm:mainresult}
  There is a one-to-one correspondence between generic holomorphic quadrics in the dual twistor space of flat space, compatible with Euclidean reality conditions, and self-dual black hole metrics.   
\end{theorem}
In particular, we show that each of the three classes of dual twistor quadrics gives rise to self-dual Taub-NUT, Eguchi-Hanson and self-dual Pleba\'nski-Demia\'nski metrics, respectively. Furthermore, we obtain as a corollary that each of these self-dual black hole metrics admits a (single) Kerr-Schild form. While Eguchi-Hanson was long known to be Kerr-Schild~\cite{Sparling:1981nk,Tod:1982mmp}, it was only recently proved that self-dual Taub-NUT has a Kerr-Schild form~\cite{Kim:2024dxo} and (to our knowledge) this is a totally new result for the case of self-dual Pleba\'nski-Demia\'nski. 

\medskip

The paper is organised as follows. In Section~\ref{sec:SDBH} we give an introduction to the self-dual black hole metrics, starting from the general Pleba\'nski-Demia\'nski family of type D solutions. While much of this exposition will be familiar to experts, we do note a previously overlooked degenerate case of the self-dual Pleba\'nski-Demia\'nski metric. Section~\ref{sec: dual twistor quadrics to SD Kerr-Schild metrics} gives the algorithm for constructing hyperk\"ahler Kerr-Schild metrics from dual twistor quadrics. We begin with an overview of the necessary concepts in twistor and dual twistor theory, before reviewing Tod's construction of hyperk\"ahler Kerr-Schild metrics from null self-dual Maxwell fields in flat space. We then show how every dual twistor quadric gives rise to such a null self-dual Maxwell field, and prove that the resulting metrics always admit Gibbons-Hawking (in addition to Kerr-Schild) forms. 

In Section~\ref{sec:DTQClass}, we classify all dual twistor quadrics, showing that they are all fall into one of three `types', which we refer to as Cases A, B and C. We then prove that these each correspond to one of the self-dual black hole metrics, with Case A corresponding to self-dual Taub-NUT, B to Eguchi-Hanson and C to self-dual Pleba\'nski-Demia\'nski. Section~\ref{sec:Discussion} concludes with some speculations about future directions, particularly with regards to our initial motivation of developing a new approach to black hole perturbation theory.


\section{Self-dual black hole metrics}
\label{sec:SDBH}

All algebraically type D solutions of Einstein's equations in Lorentzian signature with a doubly-aligned Maxwell field\footnote{It is worth noting the recent discovery of new, type D solutions of the Einstein-Maxwell equations where the Maxwell field is \emph{not} aligned with the two principal null vectors of the metric~\cite{Podolsky:2025tle,Ovcharenko:2025cpm}. These solutions can be interpreted as Kerr black holes in a uniform external magnetic field.}, with or without cosmological constant, are given by the Pleba\'nski-Demia\'nski (PD) family of metrics~\cite{Debever:1971,Plebanski:1976gy}. In its most general form, a metric in this family is specified by seven parameters whose physical meaning is not typically clear; however, in various limits these can be identified with mass, NUT charge, electric and magnetic charge, spin/twist, acceleration and cosmological constant. In various limits, the PD family reduces to many other well-known solutions (cf., \cite{Plebanski:1976gy,Griffiths:2005qp,Ovcharenko:2024yyu}), including the Kerr-Newman family of astrophysical black holes~\cite{Kerr:1963ud,Newman:1965my}, the Taub-NUT spacetime~\cite{Taub:1950ez,Newman:1963yy} and the C-metric describing a pair of causally-separated, accelerating black holes~\cite{Weyl:1917gp,Kinnersley:1970zw,Griffiths:2006tk}. 

As Lorentzian-real metrics, there are, of course, no self-dual examples of the PD family besides the trivial cases of the Minkowski, de Sitter and anti-de Sitter spacetimes. However, by complexifying the PD metrics (i.e., extending them to holomorphic, complex metrics), it is possible to obtain self-dual solutions, which can then be considered as real metrics in non-Lorentzian signature (i.e., Euclidean or split signature) by taking an appropriate real slice. As the Lorentzian PD family describes general, type D black hole metrics, it seems natural to call the resulting metrics \emph{self-dual black holes}, although clearly they will not be black hole metrics in the usual Lorentzian sense.

We will confine our attention to vacuum, asymptotically flat self-dual black hole metrics in Euclidean signature. More precisely:
\begin{defn}[Self-dual black hole]\label{SDBHdef}
A Ricci-flat, self-dual Riemannian metric (i.e., a hyperk\"ahler metric) will be referred to as a \emph{self-dual black hole} if it arises as a Euclidean-real slice of the complexification of some member of the Pleba\'nski-Demia\'nski family of metrics.
\end{defn}
In this section, we give an overview of these self-dual black holes, starting with a brief review of the general Euclidean PD family. We then discuss the three distinct cases of self-dual black holes, beginning with the most general self-dual Pleba\'nski-Demia\'nski metric followed by the more familiar Eguchi-Hanson and Taub-NUT gravitational instantons.


\subsection{Review: the Pleba\'nski-Demia\'nski metrics}

While the original PD family of metrics was written in terms of seven parameters, the Einstein-Maxwell equations actually reduce the true number of free parameters to six (cf., \cite{Griffiths:2005qp,Chen:2015vva,Araneda:2023xnv}). Setting the cosmological constant to zero, this leaves five free parameters for the asymptotically flat PD family. Needless to say, the relationship between these five parameters and the original PD parametrization is highly non-trivial.

For our purposes, the most natural form of the PD metric is~\cite{Lapedes:1980qw,Chen:2015vva}
\begin{align}
g = \frac{1}{(p-q)^2}\left[ -\mathcal{Q}\,\frac{(\d\tau-p^2\d\phi)^2}{(1-p^2q^2)} + \mathcal{P}\,\frac{(\d\phi - q^2\d\tau)^2}{(1-p^2q^2)}+(1-p^2q^2)\left(\frac{\d{p}^2}{\mathcal{P}}-\frac{\d{q}^2}{\mathcal{Q}} \right) \right], \label{PDmetric}
\end{align}
where 
\begin{align}\label{PDpolynomials}
\mathcal{P} = a_0 + a_1\, p + a_2\, p^2 + a_3\, p^3 + a_4\, p^4, \qquad 
\mathcal{Q} = a_0 + a_1\, q + a_2\, q^2 + a_3\, q^3 + a_4\, q^4,
\end{align}
and $a_0\ldots,a_4$ are free parameters. While this can be viewed as a holomorphic complex metric in the first instance, this particular form of the PD metric is naturally adapted to Euclidean signature. Indeed, when the coordinates $(\tau,\phi,p,q)$ are real with ranges such that $\mathcal{P}>0$ and $\mathcal{Q}<0$, it is easy to see that \eqref{PDmetric} is a real, Riemannian metric.  

For any choice of $a_0,\ldots,a_4$, the metric \eqref{PDmetric} is a solution to the Einstein-Maxwell equations. This solution is furthermore Ricci-flat if $a_4=a_0$, self-dual if $a_4=a_0$ and $a_3=a_1$, and flat if $a_4=a_0$ and $a_3=a_1=0$. This is further illustrated by examining the only non-trivial Weyl scalars for \eqref{PDmetric}-\eqref{PDpolynomials}, which are (see e.g., \cite[Eq. (4.24)]{Araneda:2023xnv} \footnote{Our orientation conventions are opposite to those in \cite{Araneda:2023xnv}, so $\Psi_{2}^{\pm}$ here corresponds to $\Psi_{2}^{\mp}$ in \cite{Araneda:2023xnv}.})
\begin{align}\label{Psi2PD}
 \Psi^{\pm}_{2} = -\frac{(a_3\pm a_1)}{2}\left(\frac{p-q}{1\mp pq}\right)^{3} + 
 (a_0-a_4)\left(\frac{p+q}{1\pm pq}\right)\left(\frac{p-q}{1\mp pq}\right)^{3}\,,
\end{align}
as expected for a type D metric.

For any $a_0,...,a_4$, the metric \eqref{PDmetric} is toric, conformally K\"ahler (on both orientations), has zero scalar curvature and is asymptotically locally Euclidean (ALE)~\cite{Chen:2015vva,Araneda:2023xnv}. It then has an SU$(\infty)$ Toda formulation (cf., \cite{Bakas:1989xu,Park:1989vq,Tod:1995vm}): there are coordinates $(\psi,x,y,z)$ with functions $u,W$ and a 1-form $A$ such that the geometry is described by the system
\begin{align}
 g ={}& W^{-1}\,(\d\psi+A)^2+W\left[\d{z}^2+\e^{u}\,(\d{x}^2+\d{y}^2)\right]\,, \label{Todametric} \\
 \d{A} ={}& \left[W_{x}\,\d{z} - z^2\,\partial_{z}\left(\frac{W\, \e^{u}}{z^2}\right)\d{x}\right]\wedge\d{y}\,, \label{monopoleA} \\
 0={}& W_{xx}+\partial_{z}\left[z^2\,\partial_{z}\left(\frac{W\, \e^u}{z^2}\right)\right]\,, \label{monopoleW} \\
 0 ={}& u_{xx}+(\e^{u})_{zz}\,, \label{Todaeq}
\end{align}
with subscripts denoting partial derivatives. The transformation that takes the PD metric \eqref{PDmetric} to the SU$(\infty)$ form \eqref{Todametric} is \cite{Araneda:2023xnv}
\begin{gather}
\nonumber \psi = \frac{\tau+\phi}{2}\,, \qquad 
y=-\tau+\phi\,, \qquad z = \frac{1-pq}{p-q}\,, \qquad 
\d{x}=\frac{(1-p^2)}{\mathcal{P}}\,\d p - \frac{(1-q^2)}{\mathcal{Q}}\,\d q\,, \label{TodaPD1} \\
\nonumber \frac{1}{W} = \frac{(1-q^2)^2\, \mathcal{P} - (1-p^2)^2\, \mathcal{Q}}{(p-q)^2\,(1-p^2q^2)}, \qquad 
\e^{u} = \frac{-\mathcal{P}\,\mathcal{Q}}{(p-q)^4}\,, \\
A = \frac{[(1-q^4)\,\mathcal{P}+(1-p^4)\,\mathcal{Q}]}{[(1-q^2)^2\,\mathcal{P}-(1-p^2)^2\,\mathcal{Q}]}\,\frac{(\d\phi-\d\tau)}{2}\,.
\label{TodaPD2}
\end{gather}
For the chosen orientation, the conformal K\"ahler structure is 
\begin{align}\label{CKSPD}
\omega = \frac{(\d\tau-p^2\d\phi)\wedge\d{q}+\d{p}\wedge(\d\phi-q^2\d\tau)}{(p-q)^2}\,, 
\qquad 
\Omega = \frac{p-q}{1-pq}\,,
\end{align}
meaning that the tensor field $J^{a}{}_{b}=\omega_{bc}\,g^{ca}$ is an integrable almost-complex structure, which is parallel with respect to the Levi-Civita connection of the metric $\Omega^2\, g_{ab}$.

Note that both the fundamental 2-form $\omega$ and the conformal factor $\Omega$ are, remarkably, independent of the PD parameters $a_0,...,a_4$. This implies, in particular, that the conformally K\"ahler structure remains non-trivial in both the self-dual and flat limits of PD.

Defining $\rho^2:=\e^{u}$, equation \eqref{Todaeq} implies that there is a function $Z$ defined by $Z_{z}=\rho_{x}/\rho$, $Z_{x}=-\rho\,\rho_{z}$. The pair $(\rho,\,Z)$ are called Weyl-Lewis-Papapetrou coordinates, and are given explicitly by: 
\begin{align}\label{WPPD}
\rho^2 = \frac{-\mathcal{P}\,\mathcal{Q}}{(p-q)^4}, \qquad 
Z =  -\frac{[2a_0+a_1\,(p+q)+pq\,(2 a_2 + 2a_4\, pq + a_3\,(p+q))]}{2\,(p-q)^2}\,.
\end{align}
These various structures will play an important role in our explorations of the self-dual black hole metrics lying within the PD family.


\subsection{Self-dual Pleba\'nski-Demia\'nski}
\label{sec:SDPD0}

Given the general form of the PD metric \eqref{PDmetric} with Euclidean reality conditions, the first example of a self-dual black hole -- as characterised by Definition~\ref{SDBHdef} -- is provided by simply restricting the parameters to give a vacuum, self-dual metric. As stated above, this occurs when $a_4=a_0\neq0$ and $a_3=a_1$, meaning that the functions $\mathcal{P}$, $\mathcal{Q}$ are given by
\be\label{SD-PQ}
\mathcal{P} = a_0 + a_1\, p + a_2\, p^2 + a_1\, p^3 + a_0\, p^4\,, \qquad 
\mathcal{Q} = a_0 + a_1\, q + a_2\, q^2 + a_1\, q^3 + a_0\, q^4\,.
\ee
Consequently, this \emph{self-dual Pleba\'nski-Demia\'nski} (SDPD) metric has three free parameters: $a_0,a_1,a_2$ in this description.

However, it is arguably more illuminating to consider the SDPD metric in the SU$(\infty)$ Toda formulation. In this case, a direct calculation with $a_4=a_0$, $a_3=a_1$ gives:
\begin{align}\label{PDGHform}
 g_{\mathrm{SDPD}} = \frac{(\d T+\omega)^2}{V}+V\left(\d\rho^2+\d{Z}^2+\rho^2\,\d{\varphi}^2\right),
\end{align}
where $(\rho,Z)$ are the Weyl-Lewis-Papapetrou coordinates of \eqref{WPPD} evaluated in the self-dual parametrization and 
\begin{align}\label{variablesGHPD}
T:=\ell^{-1}\,\psi\,, \quad \varphi:=\ell\, y\,, \quad \omega:=\ell^{-1}\,A\,, \quad V:=\ell^{-2}\,W\,,
\end{align}
for the quantity
\be\label{elldef}
\ell^2:=\frac{8\,a_0^2+a_1^2-4\,a_0\,a_2}{4}\,,
\ee
with the assumption that $\ell\neq0$. 

In this case, equation \eqref{monopoleA} becomes the abelian monopole equation relating $\omega$ and $V$:
$\d\omega=\star_{3}\d V$, where $\star_{3}$ is the Hodge star of the flat 3-metric $\d\rho^2+\d{Z}^2+\rho^2\d{\varphi}^2$. Equation \eqref{monopoleW} then becomes the axisymmetric Laplace equation for $V$. Consequently, we see that the self-dual limit of \eqref{PDmetric} is described by the Gibbons-Hawking ansatz with potential 
\begin{align}\label{VSDPD0}
 V_{\mathrm{SDPD}} = \frac{1}{\ell^2}\frac{(p-q)\,(1+pq)}{[(a_2+2a_0)\,(p+q)\,(1+pq)+a_1\,(1+4pq+p^2+q^2+p^2q^2)]}\,.
\end{align}
As it stands, this potential is not recognizable as a multi-centre fundamental solution of the axisymmetric Laplace equation typically encountered in Gibbons-Hawking metrics.

To cast this in a more recognizable form, one first recombines the parameters into the following combinations:
\begin{subequations}\label{bapm}
\begin{align}
 b_{\pm}^{2} :={}& \frac{1}{4}\left[ a_1^2 - 2 a_0 (a_2 + 2a_0) \pm 2 a_0  \sqrt{ (2a_0+a_2)^2 - 4a_1^2} \right], \\
 a_{\pm} :={}& \frac{a_1}{8}\left[ a_2 - 6a_0 \pm \sqrt{ (2a_0+a_2)^2 - 4a_1^2} \right], \\
Z_{\pm} :={}& \frac{1}{4}\left[ a_2-2a_0 \pm \sqrt{ (2a_0+a_2)^2 - 4a_1^2} \right].
\end{align}
\end{subequations}
These quantities can then be used in turn to define the variables
\begin{align}
 r_{\pm}^{2}:=\frac{1}{b_{\pm}^2}\frac{[a_{\pm}\,(p+q)+b_{\pm}^2\,(1+pq)]^2}{(p-q)^2}\,,
\end{align}
which can be shown to obey
\begin{align}\label{rpm}
 r_{\pm}^2 = \rho^2 + (Z - Z_{\pm})^{2}\,.
\end{align}
Thus, $r_{\pm}$ define distances with respect to the flat 3-metric $\d\rho^2+\d{Z}^2+\rho^2\d{\varphi}^2$ in \eqref{PDGHform}.

One can then check that the potential \eqref{VSDPD0} becomes
\begin{align}\label{VSDPD1}
 V_{\mathrm{SDPD}} = \frac{m_{+}}{r_{+}} + \frac{m_{-}}{r_{-}}\,, 
 \qquad \text{where} \qquad
 m_{\pm}:=\frac{\pm a_{\pm}}{b_{\pm}\,\ell^2\,\sqrt{(2a_0+a_2)^2-4a_1^2}}\,.
\end{align}
Thus, the SDPD metric, given by \eqref{PDGHform} with \eqref{VSDPD1}, corresponds to an ALE two-centred Gibbons-Hawking hyperk\"ahler metric, a rather non-trivial fact which was first observed nearly 25 years ago~\cite{Casteill:2001zk}. Observe that in this form, the three parameters of the SDPD metric are $m_{-}$, $m_+$ and the relative separation between the centres, $r_+ - r_-$.

\medskip

\paragraph{Degenerate case:} Now, this analysis has assumed the generic case in which the two centres are different. The situation in which the centres coincide can be obtained as a limit
\begin{align}\label{deglimit}
{Z}_{+} - {Z}_{-} \to 0,
\end{align}
which corresponds to $(2a_0+a_2)^2-4a_1^2 \to 0$. 
This is, in fact, a degenerate limit, requiring a separate treatment; to our knowledge, this has not been noticed before in the literature\footnote{Given a multi-centred Gibbons-Hawking metric, the procedure by which two centres coalesce is described by the phenomenon of collapsing geometries and bubbling-off ALE spaces (cf., \cite{Sun:2021}). In the SDPD case, due to the dependence of the masses on the separation distance, this description does not apply.}. 

Defining the coordinates and parameters
\begin{equation}
\begin{aligned}
& x_1 := \rho \cos\varphi\,, \qquad 
x_2 := \rho\sin\varphi\,, \qquad 
x_3 := Z-\frac{(a_2-2a_0)}{4}\,, \\
& \varepsilon:=\frac{\sqrt{(2a_0+a_2)^2-4a_1^2}}{4}\,, \qquad 
c:=\frac{(a_2-6a_0)}{4}\,,
\end{aligned}
\end{equation}
the SDPD metric becomes
\begin{subequations}\label{eps2centres}
\begin{align}
g_{\varepsilon} ={}& \frac{1}{V_{\varepsilon}}\,(\d\psi+A_{\varepsilon})^{2} + V_{\varepsilon}\,(\d{x}_1^2+\d{x}_2^2+\d{x}_3^2)\,, \label{geps2centres}  \\
V_{\varepsilon} ={}& \frac{m_{+}(\varepsilon)}{\sqrt{x_1^2+x_2^2+(x_3-\varepsilon)^2}}+\frac{m_{-}(\varepsilon)}{\sqrt{x_1^2+x_2^2+(x_3+\varepsilon)^2}}, \label{Veps2centres}
\end{align}
\end{subequations}
where the masses are
\begin{align}\label{masseps}
m_{\pm}(\varepsilon) = \frac{a_1}{8\ell^2\, b_{\pm}(\varepsilon)}\left(\frac{\varepsilon \mp c}{\varepsilon} \right)
\end{align}
with $b^2_{\pm}(\varepsilon)=\frac{a_1^2-2a_0(2a_0+a_2)}{4}\pm 2 a_0\varepsilon$, and the dependence on $\varepsilon$ of all quantities has been emphasized. The coalescing-centres limit \eqref{deglimit} now corresponds to $\varepsilon\to0$. 

A Taylor expansion of $V_{\varepsilon}$ around $\varepsilon=0$ gives
\begin{align}\label{VSDPDe}
V_{\varepsilon} = \frac{A}{\sqrt{x_1^2+x_2^2+x_3^2}}+\frac{B \, x_3}{(x_1^2+x_2^2+x_3^2)^{3/2}} + O(\varepsilon^2)
\end{align}
where 
\begin{align}
 A = \frac{a_0+a_2}{2c^{5/2}\sqrt{2a_0+a_2}}\,, \qquad 
 B = \frac{\sqrt{2a_0+a_2}}{8c^{3/2}}\,.
\end{align}
Thus, in the coalescing-centres limit \eqref{deglimit} (i.e. $\varepsilon\to 0$), the SDPD metric retains the Gibbons-Hawking form, but the potential does not correspond to a multi-centred solution anymore: expressing the limit $\varepsilon\to 0$ of \eqref{VSDPDe} in spherical coordinates gives 
\begin{align}\label{VSDPDe=0}
\lim_{\varepsilon\to 0}V_{\varepsilon} 
= \frac{A}{r} - \frac{\partial}{\partial x_{3}}\left( \frac{B}{r} \right).
\end{align}
The Laplacian of this function produces $-4\pi (A\delta^{(3)}({\bf x})+B\partial_{x_3}\delta^{(3)}({\bf x}))$ (where ${\bf x}=(x_1,x_2,x_3)$). While the singularity structure is still point-like, it is now angle-dependent.


\subsection{Eguchi-Hanson}

The standard form of the Eguchi-Hanson (EH) metric is~\cite{Eguchi:1978xp,Eguchi:1978gw} 
\begin{align}\label{EguchiHanson0}
 g_{\rm EH} = f(r)\,\frac{r^2}{4}\left(\d\psi+\cos\theta\,\d\phi\right)^2 + \frac{\d r^2}{f(r)} + \frac{r^2}{4}\left(\d\theta^2+\sin^2\theta\,\d\phi^2\right)\,,
\end{align}
where $f(r)=1-(\frac{a}{r})^4$ and $r>a$, $0\leq\psi\leq2\pi$, $(\theta,\phi)\in S^2$. As shown by Prasad~\cite{Prasad:1979kg}, the EH metric \eqref{EguchiHanson0} is isometric to an ALE two-centred Gibbons-Hawking metric with equal masses. The coordinate transformation is most easily seen starting from the Gibbons-Hawking form
\begin{align}\label{GHEH}
 g_{\mathrm{EH}} = V^{-1}\left(\d\tau+A\right)^2+V\left(\d\mathcal{X}^2+\d\mathcal{Y}^2+\d\mathcal{Z}^2\right)\,,
\end{align}
for
\be\label{EHparam}
V=\frac{1}{\mathcal{R}_{+}}+\frac{1}{\mathcal{R}_{-}}\,, \qquad A=\left(\frac{\mathcal{Z}_{+}}{\mathcal{R}_{+}}+\frac{\mathcal{Z_{-}}}{\mathcal{R}_{-}}\right)\d\tan^{-1}\left(\frac{\mathcal{Y}}{\mathcal{X}}\right)\,,
\ee
where $\mathcal{Z}_{\pm}:=\mathcal{Z}\pm\frac{a^2}{8}$ and $\mathcal{R}_{\pm}:=\sqrt{\mathcal{X}^2+\mathcal{Y}^2+\mathcal{Z}_{\pm}^2}$. It is easy to verify that these are related by the monopole equation $\d A=\star_3 \d V$, where $\star_3$ is the Hodge star of the flat metric on $\R^3$ in the Cartesian coordinates $(\mathcal{X},\mathcal{Y},\mathcal{Z})$. Following~\cite{Prasad:1979kg}, the coordinates $(\psi,r,\theta,\phi)$ are obtained by
\begin{align}
\tau = 2\phi\,, \quad
\mathcal{X} = \frac{r^2}{8}\sqrt{f(r)}\sin\theta\cos\psi\,, \quad
\mathcal{Y} = \frac{r^2}{8}\sqrt{f(r)}\sin\theta\sin\psi\,, \quad 
\mathcal{Z} = \frac{r^2}{8}\cos\theta\,.
\end{align}
Using this coordinate transformation in the Gibbons-Hawking metric \eqref{GHEH}, one obtains the original form \eqref{EguchiHanson0} of the EH metric.

\medskip

Given that the SDPD metric \eqref{PDGHform}-\eqref{VSDPD1} is an ALE two-centred Gibbons-Hawking metric with different masses, it is tempting to say that the EH solution is just a special case of the SDPD metric where the masses are equal. If this were indeed the case, then it would be obvious that EH is a self-dual black hole, but there would be no need to consider it as distinct from SDPD. Surprisingly, it is \emph{not} possible to obtain EH from SDPD while keeping the metric Euclidean-real, so while EH is indeed a self-dual black hole (being obtained from a complexification of SDPD), it is distinct from SDPD as a Riemannian metric. This follows from the observation that the Euclidean PD metric is \emph{never} regular~\cite{Araneda:2025uqo}, whereas the EH metric is (famously) a regular, ALE gravitational instanton.


\subsection{Self-dual Taub-NUT}

Within the \emph{complex} PD family of metrics, one can consider the sub-family of metrics with vanishing cosmological constant, electric and magnetic charges -- for a certain range of the remaining parameters, these metric can be understood as accelerating Taub-NUT spacetimes with rotation~\cite{Plebanski:1976gy,Griffiths:2005qp}. Upon sequentially setting the parameters corresponding to acceleration and rotation to zero, one obtains the usual Taub-NUT metric in the holomorphic category, which is a two-parameter family of metrics specified by mass $M$ and NUT charge $N$. 

This complex metric becomes (locally) vacuum self-dual when $N=-\im\,M$; in complexified Newman-Unti-Tamburino~\cite{Newman:1963yy} coordinates $(t,r,\theta,\phi)$ this is given by
\be\label{complexTN}
g_{\mathrm{SDTN}} = -\left(\frac{r-M}{r+M}\right)\left(\d t+2\im\,M\,\cos\theta\,\d\phi\right)^2+\left(\frac{r+M}{r-M}\right)+\left(r^2-M^2\right)\,\d\Omega^2\,,
\ee
with $\d\Omega^2$ the usual line element on the (complexified) round unit sphere:
\be\label{sphermet}
\d\Omega^2:=\d \theta^2+\sin^2\theta\,\d\phi^2\,.
\ee
This metric admits a Euclidean-real slice for coordinates $(\tau=\im\,t,r,\theta,\phi)$ and $M\in(0,\infty)$; after taking $r-M\to r$ the Euclidean self-dual Taub-NUT (SDTN) becomes~\cite{Hawking:1976jb} 
\be\label{SDTNmet}
g_{\mathrm{SDTN}} = V^{-1}\left(\d\tau+A\right)^2+V\left(\d r^2+r^2\,\d\Omega^2\right)\,,
\ee
with
\be\label{SDTN-GH}
V=1+\frac{2\,M}{r}\,, \qquad A=\cos\theta\,\d\phi\,.
\ee
It is easy to show that $\d A=\star_{3}\d V$, where $\star_3$ is the Hodge star on $\R^3\setminus{0}$ with metric $\d r^2+r^2\,\d\Omega^2$, so the SDTN metric is an ALF Gibbons-Hawking metric. In fact, the metric gives globally smooth gravitational instanton on $\R^4$, with the periodic identification of Euclidean time $\tau\sim\tau+8\pi\,M$ removing the singularities at $\theta=0,\pi$ which would otherwise be present.

\medskip

It is also possible to obtain the SDTN metric directly from the Euclidean PD metric, albeit via a singular limit. To see this, consider a particular case of \eqref{PDmetric}-\eqref{PDpolynomials} in which the coefficients $a_0,...,a_4$ are re-parameterised in terms of new constants $m,n,a,c$ as
\begin{align}
a_{0} = \frac{a^2-n^2}{c^4}\,, \qquad a_1=\frac{2n}{c^3}\,, \qquad a_2=\frac{-1}{c^2}\,, \qquad a_3 = \frac{2m}{c^3}\,, \qquad a_4 =  \frac{a^2-n^2}{c^4}\,,
\end{align}
and the coordinates $(\tau,\phi,p,q)$ are mapped to a new system $(\psi,\varphi,r,\theta)$ by 
\begin{align}
\psi = \frac{\tau}{c}-\frac{(a^2+n^2)}{c^3}\phi\,, \qquad \varphi = \frac{a}{c^3}\phi\,, \qquad r = \frac{c}{q}\,, \qquad \cos\theta = \frac{c p - n}{a}\,.
\end{align}
Upon taking the $c\to\infty$ limit, the PD metric \eqref{PDmetric}-\eqref{PDpolynomials} becomes
\begin{multline}\label{KTN}
 g = \frac{\Delta}{\Sigma}\left[\d\psi+(2n\,\cos\theta+a\,\sin^2\theta)\,\d\varphi\right]^2 + \frac{\sin^2\theta}{\Sigma}\left[a\,\d\psi-(r^2-a^2-n^2)\,\d\varphi\right]^2 \\
 +\Sigma\left(\frac{\d{r}^2}{\Delta}+\d\theta^2\right)\,,
\end{multline}
where $\Delta = r^2-2mr-a^2+n^2$, $\Sigma = r^2-(n-a\cos\theta)^2$. This is the three-parameter family of (Ricci-flat) Euclidean Kerr-NUT metrics, parameterised by $m,n,a$. 

From \eqref{Psi2PD}, one finds that the Weyl curvature of this metric is
\begin{align}\label{Psi2KTN}
\Psi_{2}^{\pm} = \frac{m\mp n}{[r\mp(n-a\,\cos\theta)]^{3}}\,.
\end{align}
Setting $a=0$, gives the two-parameter family of Taub-NUT metrics, with parameters $m=M$ and $n=-\im\,N$. Further setting $m=-n$, we see from \eqref{Psi2KTN} that $\Psi_2^{-}=0$, so the curvature is self-dual, and the SDTN metric is obtained. Note that while complexification of the PD metric \eqref{PDmetric} was not required to get \eqref{KTN}, the $c\to\infty$ limit is singular, insofar as it changes the asymptotic structure from ALE to ALF. 

\medskip

This establishes that the SDTN metric is a self-dual black hole metric by Definition~\ref{SDBHdef}. It should be pointed out that although SDTN is this simplest of the self-dual black hole metrics (in the sense that it is globally defined, topologically trivial and has only a single parameter), it has frequently been used as the prototype of a self-dual black hole in the literature (cf., \cite{Crawley:2023brz,Guevara:2023wlr,Adamo:2023fbj,Guevara:2024edh,Adamo:2025fqt}). This is due to the fact that the complexified SDTN metric \eqref{complexTN} also admits a real slice in split signature. This split signature SDTN metric has a null `horizon' (where the positive and negative signature components of the metric swap signs), behind which the metric can be extended to a true curvature singularity~\cite{Crawley:2023brz}. While in split signature the usual causal interpretation of black holes is not available, the presence of a horizon and singularity would seem to further justify the nomenclature of `self-dual black hole.' It would be interesting to explore whether Eguchi-Hanson or SDPD have similar features in split signature.


\section{From dual twistor quadrics to SD Kerr-Schild metrics}\label{sec: dual twistor quadrics to SD Kerr-Schild metrics}

All three classes of self-dual black hole metrics share two important properties: self-duality, and being algebraically special of type D. The former means that all self-dual black holes have a description in terms of \emph{twistor theory}, via the famous non-linear graviton construction~\cite{Penrose:1976js,Atiyah:1978wi}. In essence, this construction states that there is a one-to-one correspondence between any self-dual 4-manifold and a complex deformation of an open subset of the complex projective space $\P^3$ with some technical assumptions. However, the second property (algebraically type D) also has important consequences. 

In this section, we show that any holomorphic quadric in the dual twistor space of \emph{flat} space gives rise to a self-dual, type D metric in Kerr-Schild form, with the metric, its hyperk\"ahler and conformally K\"ahler structures and its transformation to Gibbons-Hawking form all encoded explicitly in the data of the dual twistor quadric. After a brief overview of some salient facts about twistor and dual twistor space, we lay out the construction of self-dual type D metrics from generic dual twistor quadrics. 


\subsection{Twistor and dual twistor spaces}

Let $Z^{A}=(\mu^{\dot\alpha},\lambda_{\alpha})$ be homogeneous coordinates on the complex projective space $\P^3$ and define
\be\label{twistorspace}
\PT=\left\{Z\in\P^3\,|\,\lambda_{\alpha}\neq 0\right\}\,,
\ee
to be the twistor space of complexified flat space, $\M$ (our notational conventions follow~\cite{Adamo:2017qyl}). Points $x^{\alpha\dot\alpha}\in\M$ are described by holomorphic rational curves in $\PT$ with normal bundle $\cO(1)\oplus\cO(1)$; these are given explicitly by the `incidence relations'
\be\label{ogir}
\mu^{\dot\alpha}=x^{\alpha\dot\alpha}\,\lambda_{\alpha}\,.
\ee
To obtain Euclidean $\R^4\subset\M$, one imposes appropriate reality conditions in twistor space. In particular, defining the `quaternionic conjugation' operation
\be\label{EuclidConj}
\hat{Z}^{A}:=\left(-\overline{\mu^{\dot{1}}},\,\overline{\mu^{\dot{0}}},\,-\overline{\lambda_{1}},\,\overline{\lambda_{0}}\right)\,,
\ee
the twistor lines corresponding to points $x\in\R^{4}$ are those which are fixed by this operation (cf., \cite{Atiyah:1978wi,Woodhouse:1985id}). 

The non-linear graviton construction states that any self-dual metric can be encoded by a complex deformation of some open subset of $\PT$~\cite{Penrose:1976js,Atiyah:1978wi}; in practical terms, this boils down to a Hamiltonian deformation of the complex structure
\be\label{PTdeform}
\dbar \rightarrow\dbar+\frac{\partial h}{\partial\mu_{\dot\alpha}}\,\frac{\partial}{\partial\mu^{\dot\alpha}}\,,
\ee
for $h\in\Omega^{0,1}(\PT,\cO(2))$ obeying the integrability condition
\be\label{PTdeform2}
\dbar h+\frac{1}{2}\,\frac{\partial h}{\partial\mu_{\dot\alpha}}\wedge\frac{\partial h}{\partial\mu^{\dot\alpha}}=0\,.
\ee
Points in the associated self-dual space are given by rational curves, holomorphic with respect to the complex structure \eqref{PTdeform}. These are defined by sections $\mu^{\dot\alpha}=F^{\dot\alpha}(x,\lambda)$, homogeneous of weight $+1$ which are solutions to the PDE
\be\label{holcurves}
\dbar F^{\dot\alpha}=\frac{\partial h}{\partial\mu_{\dot\alpha}}\,, 
\ee
on the Riemann sphere.

While theorems of Kodaira~\cite{Kodaira:1962,Kodaira:1963} guarantee the existence of a 4-parameter family of solutions to this equation (for sufficiently `small' data $h$), determining the explicit form of these holomorphic curves is not necessarily easy. However, reconstructing the self-dual metric associated to the twistor space requires explicit knowledge of the holomorphic curves (cf., \cite{Gindikin:1986}). This is a typical feature of integrable systems, where the power of integrability is gained by solving an `associated linear problem.'

\medskip

The second unifying feature of self-dual black holes -- namely, that they are algebraically special of type D -- enables an alternative description. Remarkably, this description is based on \emph{dual twistor space}, rather than the twistor space associated with self-duality. The possibility of describing self-dual type D metrics with dual twistor theory was noted long ago by Haslehurst and Penrose~\cite{Haslehurst:1992} and by Woodhouse~\cite{Woodhouse:1992}, building on a variety of earlier work (cf., ~\cite{Tod:1979tt,Penrose:1983,Ward:1983,Penrose:1984}). While these constructions essentially used a curved dual twistor space, our version of the correspondence will make use only of simple geometric objects in the dual twistor space of \emph{flat space}, from which we will recover all self-dual black hole metrics, along with detailed information about their hyperk\"ahler geometry.

If $\PT$ given by \eqref{twistorspace} is the twistor space of flat space, then the \emph{dual twistor space} is simply the projective dual:
\be\label{dualtwistorspace}
\PT^{*}=\left\{W_{A}=(\pi_{\dot\alpha},\,\omega^{\alpha})\,|\,\pi_{\dot\alpha}\neq 0\right\}\,,
\ee
where $W_{A}$ serve as the homogeneous coordinates on dual twistor space. This dual twistor space is related to flat space by incidence relations
\be\label{dIRs}
\omega^{\alpha}=x^{\alpha\dot\alpha}\,\pi_{\dot\alpha}\,,
\ee
which indicate that a point $x^{\alpha\dot\alpha}$ in flat space corresponds to a linearly embedded, holomorphic Riemann sphere -- or dual twistor line -- in $\PT^*$. Picking out the Euclidean-real $\R^4$ in $\M$ again induces a real structure aking to \eqref{EuclidConj}
\be\label{dtReality}
\hat{W}_{A}=\left(-\overline{\pi_{\dot{1}}},\,\overline{\pi_{\dot{0}}},\,-\overline{\omega^{1}},\,\overline{\omega^0}\right)\,,
\ee
which preserves those dual twistor lines corresponding to points in $\R^4$.

One can consider complex deformations of dual twistor space, with some technical assumptions mirroring those of the non-linear graviton construction, and this gives rise to curved, \emph{anti}-self-dual 4-manifolds. As such, it may seem un-natural to expect dual twistor space to have anything to do with self-dual black holes. However, we will see that all self-dual black hole metric arise from a simple geometric construction in \emph{flat} dual twistor space, $\PT^*$. 

While we will not need the complex deformation theory usually associated to the non-linear graviton construction (in either twistor or dual twistor space), there is one piece of standard twistor machinery which does play an important role in our construction. This is the \emph{Penrose transform}, a mechanism for solving zero-rest-mass (z.r.m.) equations in flat space from analytic data in twistor space or dual twistor space~\cite{Penrose:1969ae,Eastwood:1981jy}. 

The particular version of the Penrose transform needed for our construction is the following:
\be\label{PTrans0}
\left\{\mbox{z.r.m. fields of helicity }h\geq 0 \mbox{ on flat space}\right\}\cong H^1(\PT^*,\cO(-2-2h))\,.
\ee
Here, a z.r.m. field of helicity $h\geq0$ is a totally symmetric, valence $2h$ dotted spinor field $\varphi_{\dot\alpha_1\cdots\dot\alpha_{2h}}(x)$ which obeys
\be\label{zrm1}
\nabla^{\alpha\dot\alpha_1}\varphi_{\dot\alpha_1\cdots\dot\alpha_{2h}}=0\,.
\ee
The statement of \eqref{PTrans0} is that every solution of these equations arises from sheaf cohomology of dual twistor space, and conversely, every cohomology class gives rise to a solution. This can be understood either in terms of \v{C}ech or Dolbeault cohomology: let $f\in\check{H}^1(\PT^*,\cO(-2-2h))$ be a \v{C}ech cohomology representative and $\mathsf{f}\in H^{0,1}(\PT^*,\cO(-2-2h))$ be the corresponding Dolbeault cohomology representative (under the \v{C}ech-Dolbeault isomorphism). Then the z.r.m. field associated to these cohomology representatives is constructed by the integral formulae
\be\label{PTintforms}
\varphi_{\dot\alpha_1\cdots\dot\alpha_{2h}}(x)=\frac{1}{2\pi\im}\oint_{\Gamma\subset X} \D\pi\,\pi_{\dot\alpha_1}\cdots\pi_{\dot\alpha_{2h}}\,f|_{X}=\int_{X}\D\pi\wedge\pi_{\dot\alpha_1}\cdots\pi_{\dot\alpha_{2h}}\,\mathsf{f}|_{X}\,,
\ee
where $X\cong\P^1$ is the dual twistor line corresponding to $x^{\alpha\dot\alpha}$ defined by \eqref{dIRs} and $f|_X$, $\mathsf{f}|_{X}$ denote the pullbacks of the cohomology representatives to this line. In the \v{C}ech formula, the integral is over a real contour $\Gamma$ on the dual twistor line which separates the poles of $f|_X$ on the Riemann sphere.


\subsection{From self-dual null Maxwell fields to self-dual Kerr-Schild metrics}

The vacuum self-duality, or hyperk\"ahler, condition on a metric enables descriptions which do not make any explicit reference to twistor theory. Among these, perhaps the most famous is Pleba\'nski's `heavenly' construction, which provides a local description of any (complex) hyperk\"ahler metric in terms of scalar potentials which solve certain non-linear PDEs~\cite{Plebanski:1975wn}. One of these scalar potentials (often called the first Pleba\'nski scalar) is the K\"ahler potential for the hyperk\"ahler metric in a particular complex structure. The second Pleba\'nski scalar, on the other hand, expresses the hyperk\"ahler metric as a finite deformation of the flat hyperk\"ahler metric on $\C^4$. It is this second Pleba\'nski scalar which will be of use in this paper.

Let $M$ be a smooth 4-manifold with local coordinates $x^{a}=(u,v,w,\tilde{w})$ and a scalar function $\Theta=\Theta(u,v,w,\tilde{w})$. If $\Theta$ solves the so-called \emph{second heavenly equation}
\be\label{2HEq}
\Theta_{uv}-\Theta_{w\tilde{w}}+\Theta_{vv}\Theta_{ww}-\Theta_{vw}^2=0\,,
\ee
where subscripts denote partial derivatives, then Pleba\'nski showed~\cite{Plebanski:1975wn} that the tensor field
\be\label{2HMet}
g_{ab}\,\d{x}^{a}\,\d{x}^{b}:=2\left(\d{u}\,\d{v}-\d{w}\d\tilde{w}\right)-2\left(\Theta_{ww}\,\d{u}^2+2\,\Theta_{wv}\,\d{u}\,\d\tilde{w}+\Theta_{vv}\,\d\tilde{w}^2\right)\,,
\ee
is a complex, Ricci-flat metric on $M$, whose Weyl tensor is self-dual with respect to the orientation $\d{u}\wedge\d{v}\wedge\d{w}\wedge\d\tilde{w}$. In other words, the metric \eqref{2HMet} is automatically hyperk\"ahler by virtue of the heavenly equation \eqref{2HEq}. Furthermore, every hyperk\"ahler metric on $M$ admits a description in terms of such a potential $\Theta$ locally.

This construction can be written in terms of abstract spinor indices by making use of a covariantly constant spinor field $o_{\alpha}$ on flat space, obeying $\nabla_{\alpha\dot\alpha}o_{\beta}=0$, where $\nabla_{\alpha\dot\alpha}$ is the covariant derivative of the flat metric. Denoting
\be\label{tildederiv}
\tilde\nabla_{\dot\alpha}:=o^{\alpha}\,\nabla_{\alpha\dot\alpha}\,,
\ee
the second Pleba\'nski form of the metric becomes
\be\label{2HMet2}
g_{ab}=\eta_{ab} - 2\,o_{\alpha}\,o_{\beta}\,\tilde\nabla_{\dot\alpha}\tilde\nabla_{\dot\beta}\Theta\,,
\ee
for $\eta_{ab}$ the flat metric. Similarly, the second heavenly equation \eqref{2HEq} becomes 
\be\label{2HEq2}
\Box\Theta + \left(\tilde\nabla_{\dot\alpha}\tilde\nabla_{\dot\beta}\Theta\right)\left(\tilde\nabla^{\dot\alpha}\tilde\nabla^{\dot\beta}\Theta\right)=0\,,
\ee
where $\Box=\nabla^{\alpha\dot\alpha}\nabla_{\alpha\dot\alpha}$ is the flat space wave operator. As expected, one can show that in addition to being Ricci flat, the spinorial Weyl tensor components of the metric \eqref{2HMet2} obey $\Psi_{\alpha\beta\gamma\delta}=0$, $\tilde\Psi_{\dot\alpha\dot\beta\dot\gamma\dot\delta}\neq0$ -- that is, the metric is vacuum self-dual, or hyperk\"ahler.

\medskip

So, given a solution of the second heavenly equation \eqref{2HEq} or \eqref{2HEq2}, one automatically obtains a vacuum self-dual complex metric of the form \eqref{2HMet} or \eqref{2HMet2}, respectively. Of course, the second heavenly equation is a non-linear PDE, solutions to which should not (na\"ively, at least) be easy to find. A remarkable construction due to Tod~\cite{Tod:1982mmp} gives a way to generate solutions of the second heavenly equation directly from a null, self-dual Maxwell field in flat space. The resulting metric is then not only self-dual, but also Kerr-Schild.

A null self-dual Maxwell field is a complex abelian gauge potential $A_{\alpha\dot\alpha}$ in flat space whose field strength obeys
\be\label{SDMax1}
F_{\alpha\dot\alpha\beta\dot\beta}=\epsilon_{\alpha\beta}\,\varphi_{\dot\alpha\dot\beta}\,, \qquad \nabla^{\alpha\dot\alpha}\varphi_{\dot\alpha\dot\beta}=0\,, \qquad \varphi^{\dot\alpha\dot\beta}\,\varphi_{\dot\alpha\dot\beta}=0\,.
\ee
The second equation here, which is simply the helicity +1 z.r.m. equation, follows as a consequence of the Bianchi identity for the field strength. 

Let $\{o_{\alpha},\iota_{\alpha}\}$ be a constant dyad for the undotted spinors, with $o_{\alpha}$ the same as the spinor used to define the differential operator $\tilde{\nabla}_{\dot\alpha}=o^{\alpha}\nabla_{\alpha\dot\alpha}$ in \eqref{tildederiv} above. Without loss of generality, this constant dyad can be normalised so that $\iota^{\alpha}\,o_{\alpha}=1$. Contracting the z.r.m. for $\varphi_{\dot\alpha\dot\beta}$ with $o_{\alpha}$ immediately gives
\be\label{SDMax2}
\tilde{\nabla}^{\dot\alpha}\varphi_{\dot\alpha\dot\beta}=0\,,
\ee
which implies that there must exist, locally, a scalar function $\Theta(x^a)$ such that
\be\label{SDMaxPotential}
\varphi_{\dot\alpha\dot\beta}=\tilde{\nabla}_{\dot\alpha}\tilde{\nabla}_{\dot\beta}\Theta\,.
\ee
The justification for denoting this scalar with the same symbol as the second Pleba\'nski scalar will soon become apparent.

Observe that this does \emph{not} uniquely determine the potential $\Theta$. Indeed, for any function $f(x^a)$ satisfying
\be\label{Thetashift1}
\tilde{\nabla}_{\dot\alpha}\tilde{\nabla}_{\dot\beta}f=0\,,
\ee
it follows that the shifted potential
\be\label{Thetashift2}
\Theta\rightarrow \Theta':=\Theta+f\,,
\ee
will still produce the same null SD Maxwell field through \eqref{SDMaxPotential}. The implications of this fact are easiest understood by decomposing the coordinates $x^a$ with respect to the spinor dyad as
\be\label{coord-dyad-decomp}
x^{\dot\alpha}:= o^{\alpha}\,x_{\alpha}{}^{\dot\alpha}\,, \qquad \tilde{x}^{\dot\alpha}:=\iota^{\alpha}\,x_{\alpha}{}^{\dot\alpha}\,,
\ee
so that $\tilde{\nabla}_{\dot\alpha}\tilde{x}^{\dot\beta}=\delta_{\dot\alpha}^{\dot\beta}$. The ambiguity \eqref{Thetashift1} -- \eqref{Thetashift2} in the definition of the scalar potential $\Theta$ is then seen to amount to three unspecified functions of two variables, as
\be\label{Thetashift3}
f(x^a)=\tilde{x}^{\dot\beta}\,g_{\dot\beta}(x^{\dot\alpha})+h(x^{\dot\alpha})\,,
\ee
for any functions $g_{\dot\beta}$, $h$ of $x^{\dot\alpha}$.

Thus, given any potential $\Theta$ for $\varphi_{\dot\alpha\dot\beta}$, we can consider the alternative potential $\Theta'=\Theta+f$ for any $f$ of the form \eqref{Thetashift3}. Observing that
\be\label{Thetashift4}
\Box\Theta'=\Box\Theta+\Box f=\Box\Theta+\iota_{\alpha}\,\nabla^{\alpha\dot\beta}g_{\dot\beta}\,,
\ee
it is clear that the functions $g_{\dot\beta}(x^{\dot\alpha})$ can always be chosen so that $\Box\Theta'=0$. This choice of `gauge' for the potential $\Theta$ is consistent with Maxwell's equations, as the z.r.m. equation for $\varphi_{\dot\alpha\dot\beta}$ contracted with $\iota_{\alpha}$ is
\be\label{Thetashift5}
\tilde{\nabla}_{\dot\beta}\,\Box\Theta=0\,,
\ee
which holds, in particular, when $\Box\Theta=0$.

\medskip

So, appropriately using the freedom \eqref{Thetashift1} combined with the null condition \eqref{SDMax1} implies that $\Theta$ obeys
\be\label{SDPotEOMs}
\Box\Theta=0=\left(\tilde\nabla_{\dot\alpha}\tilde\nabla_{\dot\beta}\Theta\right)\left(\tilde\nabla^{\dot\alpha}\tilde\nabla^{\dot\beta}\Theta\right)\,.
\ee
Thus, $\Theta$ is a solution of the second heavenly equation \eqref{2HEq2} by virtue of separately setting each term in the equation to zero. In other words, every null self-dual Maxwell field in flat space defines a solution of the second heavenly equation, and therefore a vacuum self-dual metric of the form \eqref{2HMet2}.

In particular, the null self-dual Maxwell field gives the self-dual metric
\be\label{SDMaxMet1}
g_{ab}=\epsilon_{\alpha\beta}\,\epsilon_{\dot\alpha\dot\beta}-2\,o_{\alpha}\,o_{\beta}\,\varphi_{\dot\alpha\dot\beta}\,.
\ee
However, the fact that $\varphi_{\dot\alpha\dot\beta}$ is null means that it has a single, degenerate principal spinor, say $\alpha_{\dot\alpha}$, and can be written as
\be\label{SDMaxspinordecomp}
\varphi_{\dot\alpha\dot\beta}=\phi\,\alpha_{\dot\alpha}\,\alpha_{\dot\beta}\,,
\ee
for $\phi$ some overall scalar field. In particular, this means that the non-trivial part of the metric \eqref{SDMaxMet1}
\be\label{SDMaxMet2}
-2\,o_{\alpha}\,o_{\beta}\,\varphi_{\dot\alpha\dot\beta}=\kappa\,\phi\,o_{\alpha}\,o_{\beta}\,\alpha_{\dot\alpha}\,\alpha_{\dot\beta}\equiv \kappa\,\phi\,k_{a}\,k_{b}\,,
\ee
where $\kappa$ is a numerical constant (which can always be used to absorb the factor of $-2$ arising from our conventions so far) and $k_a$ is the null vector
\be\label{KSvect}
k_{\alpha\dot\alpha}:=o_{\alpha}\,\alpha_{\dot\alpha}\,.
\ee
This is precisely the condition for the metric \eqref{SDMaxMet1} to be \emph{Kerr-Schild}~\cite{Kerr:1965vyg,Debney:1969zz}.

This line of reasoning can be summarized as the following:
\begin{thm}[Tod~\cite{Tod:1982mmp}]\label{thm:Tod}
Let $\varphi_{\dot\alpha\dot\beta}=\kappa\,\phi\,\alpha_{\dot\alpha}\,\alpha_{\dot\beta}$ be a null, self-dual Maxwell field in flat space. Then there exists a scalar potential $\Theta$ and constant spinor $o_{\alpha}$ such that $\varphi_{\dot\alpha\dot\beta}=o^{\alpha}\,o^{\beta}\,\nabla_{\alpha\dot\alpha}\nabla_{\beta\dot\beta}\Theta$ and $\Box\Theta=0$, and the metric
\be\label{SDMaxMet3}
\d s^2=\d x^{\alpha\dot\alpha}\,\d x^{\alpha\dot\alpha}+o_{\alpha}\,o_{\beta}\,\varphi_{\dot\alpha\dot\beta}\,\d x^{\alpha\dot\alpha}\,\d x^{\beta\dot\beta}\,,
\ee
is hyperk\"ahler and Kerr-Schild.
\end{thm}
In other words, every null SD Maxwell field in flat space gives rise to a vacuum, SD Kerr-Schild metric.


\subsection{Self-dual null Maxwell fields from dual twistor quadrics}

Now, it can be seen see that any holomorphic quadric in the dual twistor space of flat space, $\PT^*$, gives rise to a null SD Maxwell field, and thus to a SD Kerr-Schild metric via Tod's construction.
\begin{defn}[Dual twistor quadric]
A \emph{dual twistor quadric} is a holomorphic, non-degenerate quadric hypersurface in $\PT^*$
\be\label{DTQ1}
\left\{W\in\PT^*\,|\,Q(W)=0\right\}\,, \qquad \mbox{ where }\: Q(W)=Q^{AB}\,W_{A}\,W_{B}\,,
\ee
for $Q^{AB}$ a symmetric $4\times 4$ matrix which is not of the form $Q^{AB}=A^{(A}\,B^{B)}$.
\end{defn}
This means that any dual twistor quadric corresponds to a matrix of the form
\begin{align}\label{quadric}
 Q^{AB} = 
 \left( \begin{matrix} 
 c^{\dot\alpha\dot\beta} & b^{\dot\alpha}{}_{\beta} \\
 b_{\alpha}{}^{\dot\beta} & a_{\alpha\beta}
 \end{matrix} \right)\,,
\end{align}
for some symmetric rank-2 spinors $a_{\alpha\beta}$, $b^{\dot\alpha}{}_{\beta}$, $c^{\dot\alpha\dot\beta}$. 

Armed with this definition, several important facts follow, several of which were established previously in somewhat different contexts (cf., \cite{Tod:1979tt,Araneda:2023qio}):
\begin{proposition}\label{prop:quadric}
Consider a dual twistor quadric, defined by $Q(W)$ of the form \eqref{quadric}; this defines the following geometric structures: 
\begin{enumerate}
\item\label{item:KS} The spinor field:
\begin{align}
K^{\dot\alpha\dot\beta} = a_{\alpha\beta}\,x^{\alpha\dot\alpha}\,x^{\beta\dot\beta}+2\, b_{\beta}{}^{(\dot\alpha}\,x^{|\beta|\dot\beta)}+c^{\dot\alpha\dot\beta}\,,\label{Killingspinor}
\end{align}
is a valence-2 Killing spinor, obeying $\nabla^{\gamma(\dot\gamma}K^{\dot\alpha\dot\beta)}=0$.

\item\label{item:1stKVF} Let $\xi_{\alpha\dot\alpha}:=\frac{2}{3}\nabla_{\alpha\dot\beta}K^{\dot\beta}{}_{\dot\alpha}$.
Then $\xi^{a}$ is an ASD Killing vector (i.e., $\xi^a$ is Killing and $\nabla_{\alpha\dot\alpha}\xi_{\beta\dot\beta}\propto \epsilon_{\dot\alpha\dot\beta}$).

\item\label{item:KT} The symmetric tensor field 
\be\label{Killing2Tensor}
H_{ab}:=K_{\dot\alpha\dot\beta}\,a_{\alpha\beta}-\frac{\xi^c\xi_c}{8}\,\epsilon_{\alpha\beta}\,\epsilon_{\dot\alpha\dot\beta}\,,
\ee
is a rank-2 Killing tensor, $\nabla_{(a}H_{bc)}=0$.

\item\label{item:2ndKVF} The vector field $t_{a}:=H_{ab}\,\xi^{b}$ is a SD Killing vector commuting with $\xi^{a}$.
\end{enumerate}
\end{proposition}

\proof Observe that the spinor field $K^{\dot\alpha\dot\beta}$ defined by \eqref{Killingspinor} arises from the quadric $Q$ as
\be\label{KSpinQ}
Q(W)|_{X}=K^{\dot\alpha\dot\beta}\,\pi_{\dot\alpha}\,\pi_{\dot\beta}\,,
\ee
upon using the incidence relations \eqref{dIRs}. As $\pi_{\dot\gamma}\nabla^{\gamma\dot\gamma}Q|_{X}=0$, it immediately follows that $\nabla^{\gamma(\dot\gamma}K^{\dot\alpha\dot\beta)}=0$, establishing (1.). Now, let 
\be\label{ASDKill1}
\xi_{\alpha\dot\alpha}=\frac{2}{3}\,\nabla_{\alpha\dot\beta}K^{\dot\beta}{}_{\dot\alpha}=2\left(b_{\alpha\dot\alpha}+a_{\alpha\gamma}\,x^{\gamma}{}_{\dot\alpha}\right)\,.
\ee
It immediately follows that
\be\label{ASDKill2}
\nabla_{\alpha\dot\alpha}\xi_{\beta\dot\beta}=2\,a_{\alpha\beta}\,\epsilon_{\dot\alpha\dot\beta}\,,
\ee
which is purely ASD and implies that $\nabla_{(a}\xi_{b)}=0$, establishing (2.).

From \eqref{ASDKill2}, we have 
\begin{align}
a_{\alpha\beta}\,\xi^{\beta}_{\dot\alpha} = -\frac{1}{4}\,\nabla_{\alpha\dot\alpha}(\xi_{b}\xi^{b})\,.
\end{align}
Then
\begin{equation}\label{KTens1}
\begin{split}
\nabla_{a}H_{bc} &= \epsilon_{\dot\alpha(\dot\beta}\,\xi_{\dot\gamma)\alpha}\,a_{\beta\gamma} - \frac{1}{8}\,\nabla_{a}(\xi_{d}\xi^{d})\,\epsilon_{\beta\gamma}\,\epsilon_{\dot\beta\dot\gamma} \\
&=\frac{1}{2}\left(\epsilon_{\dot\alpha\dot\beta}\,\xi_{\dot\gamma\alpha}\,a_{\beta\gamma}+\epsilon_{\dot\alpha\dot\gamma}\,\xi_{\dot\beta\alpha}\,a_{\beta\gamma}+a_{\alpha\delta}\,\xi^{\delta}_{\dot\alpha}\,\epsilon_{\beta\gamma}\,\epsilon_{\dot\beta\dot\gamma}\right)\,
\end{split}
\end{equation}
from which it follows that $\nabla_{(a}H_{bc)}=0$, establishing (3.). Finally, observe that \eqref{KTens1} implies that 
\be\label{KTens2}
\xi^{c}\,\nabla_{c}H_{ab}=a_{\alpha\beta}\,\xi^{\gamma}{}_{(\dot\alpha}\,\xi_{|\gamma|\dot\beta)} + \frac{1}{2}\epsilon_{\dot\alpha\dot\beta}\,\epsilon_{\alpha\beta}\,\xi^{\delta}{}_{\dot\gamma}\,\xi^{\gamma\dot\gamma}\,a_{\gamma\delta}=0\,,
\ee
and a direct calculation shows that $\cL_{\xi}H_{ab}=0$. These facts, along with $\nabla_{(a}H_{bc)}=0$ then imply that $\nabla_{(a}t_{b)}=0$, and that $[t,\xi]=0$. Finally, direct calculation verifies that $\nabla_{a}t_{b}\propto\epsilon_{\alpha\beta}$, so that $t^a$ is a SD Killing vector. \qed

\medskip

Note that the assumption of non-degeneracy for a dual twistor quadric implies that the associated Killing spinor $K^{\dot\alpha\dot\beta}$ is non-null; this means that there must exist spinors $\alpha_{\dot\alpha}$, $\beta_{\dot\alpha}$ obeying $\alpha^{\dot\alpha}\,\beta_{\dot\alpha}\neq0$ such that
\be\label{KillSpinDecomp}
K^{\dot\alpha\dot\beta}=\alpha^{(\dot\alpha}\,\beta^{\dot\beta)}\,.
\ee
Armed with this decomposition and the geometric objects defined by the dual twistor quadric through Proposition~\ref{prop:quadric}, we have the following key result:

\begin{proposition}\label{prop:QuadMax}
Consider any dual twistor quadric defined by $Q(W)$ which is generic in the sense that
\be\label{genericQuad}
o^{\alpha}\,\frac{\partial Q}{\partial\omega^{\alpha}}\neq 0\,,
\ee
and let $K^{\alpha\dot\alpha}$, $\xi^{\alpha\dot\alpha}$ be the associated Killing spinor and ASD Killing vector, respectively. Then the twistor function
\begin{align}\label{twistorfunction}
f(W):=-\im\,\kappa\left[ Q(W)\left(o^{\alpha}\,\frac{\partial Q}{\partial\omega^{\alpha}}\right)^2 \right]^{-1}\,,
\end{align}
is a representative of a cohomology class in $\check{H}(\PT^*,\cO(-4))$ whose Penrose transform is the null SD Maxwell field
\begin{align}\label{maxwellfield}
 \varphi_{\dot\alpha\dot\beta} = \kappa\,\phi \, \alpha_{\dot\alpha}\,\alpha_{\dot\beta}\,, 
 \qquad \phi:=\frac{-\im}{(\alpha_{\dot\alpha}\,\beta^{\dot\alpha})\,(k_{a}\,\xi^{a})^2}\,, 
 \qquad k_{\alpha\dot\alpha}:=o_{\alpha}\,\alpha_{\dot\beta}\,.
 \end{align}
\end{proposition}

\proof The definition of the quadric function $Q(W)=Q^{AB}\,W_{A}\,W_{B}$, along with the genericity assumption \eqref{genericQuad}, ensures that $f(W)$ defined by \eqref{twistorfunction} is valued in $\cO(-4)$ on $\PT^*$. That $f$ defines a \v{C}ech cohomology class in $\check{H}^1(\PT^*,\cO(-4))$ also follows by standard arguments. By the dual twistor Penrose transform, this means that $f$ defines a SD Maxwell field by the integral formula \eqref{PTintforms}:
\be\label{PTMax1}
\varphi_{\dot\alpha\dot\beta}(x)=-\frac{\kappa}{2\pi}\,\oint_{\Gamma\subset X}\D\pi\,\left.\frac{\pi_{\dot\alpha}\,\pi_{\dot\beta}}{Q\,(o^{\alpha}\,\frac{\partial Q}{\partial\omega^{\alpha}})^2}\right|_{X}\,.
\ee
Using \eqref{KillSpinDecomp}, it follows that
\be\label{PTMax2}
Q|_{X}=K^{\dot\alpha\dot\beta}\,\pi_{\dot\alpha}\,\pi_{\dot\beta}=(\alpha^{\dot\alpha}\,\pi_{\dot\alpha})\,(\beta^{\dot\beta}\,\pi_{\dot\beta})\,,
\ee
so we will take the contour $\Gamma\subset X$ to be a small circle around the point where $\alpha^{\dot\alpha}\pi_{\dot\alpha}=0$. Also, it follows that
\be\label{PTMax3}
\left.o^{\alpha}\,\frac{\partial Q}{\partial\omega^\alpha}\right|_{X}=2\left(a_{\alpha\beta}\,x^{\beta\dot\beta}+b^{\dot\beta}{}_{\alpha}\right) o^{\alpha}\,\pi_{\dot\beta}=-o_{\alpha}\,\pi_{\dot\alpha}\,\xi^{\alpha\dot\alpha}\,,
\ee
upon using the definitions from Proposition~\ref{prop:quadric}.

Evaluating the contour integral in \eqref{PTMax1} is then a straightforward residue calculation:
\be\label{PTMax4}
\begin{split}
\varphi_{\dot\alpha\dot\beta}(x)&=-\frac{\kappa}{2\pi}\,\oint_{\Gamma}\D\pi\,\frac{\pi_{\dot\alpha}\,\pi_{\dot\beta}}{(\alpha^{\dot\alpha}\,\pi_{\dot\alpha})\,(\beta^{\dot\beta}\,\pi_{\dot\beta})\,(o_{\gamma}\,\pi_{\dot\gamma}\,\xi^{\gamma\dot\gamma})^2}\\
 &=-\frac{\im\,\kappa\,\alpha_{\dot\alpha}\,\alpha_{\dot\beta}}{(\beta^{\dot\gamma}\,\alpha_{\dot\gamma})\,(k_a\,\xi^a)^2}\,,
\end{split}
\ee
for $k_a=o_{\alpha}\alpha_{\dot\alpha}$, as claimed. \qed

\medskip

There are a few remarks which are in order. Firstly, observe that the principal spinors $\alpha_{\dot\alpha}$ and $\beta_{\dot\alpha}$
are only defined up to $\alpha^{\dot\alpha}\to \nu\, \alpha^{\dot\alpha}$, $\beta^{\dot\alpha}\to \nu^{-1}\,\beta^{\dot\alpha}$ for any $\nu\in\C^*$, which leaves the Killing spinor $K^{\dot\alpha\dot\beta}$ invariant. Furthermore, the resulting null SD Maxwell field \eqref{maxwellfield} is also invariant under these rescalings. This is made slightly more explicit by re-writing the Maxwell field \eqref{maxwellfield} as
\begin{align}\label{maxwellfield2}
\varphi_{\dot\alpha\dot\beta} = \frac{\kappa}{\sqrt{2\,K^{\dot\gamma\dot\delta}\,K_{\dot\gamma\dot\delta}}}\,\frac{ \alpha_{\dot\alpha}\,\alpha_{\dot\beta}}{(o_{\alpha}\,\alpha_{\dot\alpha}\,\xi^{\alpha\dot\alpha})^2}\,,
\end{align}
having used the identity $\im\,\alpha_{\dot\alpha}\beta^{\dot\alpha}=(2K^{\dot\gamma\dot\delta}K_{\dot\gamma\dot\delta})^{1/2}$.

To compute $\alpha^{\dot\alpha}$, $\beta^{\dot\alpha}$ directly from a given dual twistor quadric, one writes
\be\label{prinspin1}
K^{\dot\alpha\dot\beta}\,\pi_{\dot\alpha}\,\pi_{\dot\beta}=K^{\dot{0}\dot{0}}\,(\pi_{\dot{0}})^2+2\,K^{\dot{0}\dot{1}}\,\pi_{\dot{0}}\,\pi_{\dot{1}}+K^{\dot{1}\dot{1}}\,(\pi_{\dot{1}})^2=(\pi_{\dot{0}})^2\, K^{\dot{1}\dot{1}}\,(\zeta-\zeta_{-})\,(\zeta-\zeta_{+})\,,
\ee
where $\zeta=\frac{\pi_{\dot{1}}}{\pi_{\dot{0}}}$ and 
\begin{align}\label{roots}
\zeta_{\pm} := \frac{-K^{\dot{0}\dot{1}} \pm \sqrt{(K^{\dot{0}\dot{1}})^2-K^{\dot{0}\dot{0}}K^{\dot{1}\dot{1}}}}{K^{\dot{1}\dot{1}}}\,.
\end{align}
One can then check that $K^{\dot\alpha\dot\beta}=\alpha^{(\dot\alpha}\,\beta^{\dot\beta)}$ for
\be\label{prinspin2}
\alpha^{\dot\alpha}=\sqrt{K^{\dot{1}\dot{1}}}\left(o^{\dot\alpha}+\zeta_{-}\,\iota^{\dot\alpha}\right)\,, \qquad \beta^{\dot\alpha}=\sqrt{K^{\dot{1}\dot{1}}}\left(o^{\dot\alpha}+\zeta_{+}\,\iota^{\dot\alpha}\right)\,,
\ee
giving the principal spinors directly in terms of the quadric data.


\subsection{Gibbons-Hawking metrics from dual twistor quadrics}

Having established that any dual twistor quadric produces a null, SD Maxwell field $\varphi_{\dot\alpha\dot\beta}$ of the form \eqref{maxwellfield}, it follows from Tod's Theorem~\ref{thm:Tod} that any dual twistor quadric will similarly give rise to a complex hyperk\"ahler metric of Kerr-Schild form \eqref{SDMaxMet3}. To better understand the structure of this new metric, first recall that, given an orientable manifold, a hyperk\"ahler structure is equivalent to a set of three linearly independent ASD, closed 2-forms. For $\mathbb{M}$, the constant spin-frame $o_{\alpha},\iota_{\alpha}$ defines a hyperk\"ahler structure via 
\begin{align}\label{flatHK}
\omega^{1}_{ab}=\i(o_{\alpha}o_{\beta} - \iota_{\alpha}\iota_{\beta})\,\epsilon_{\dot\alpha\dot\beta}\,, \qquad
\omega^{2}_{ab}=(o_{\alpha}o_{\beta} + \iota_{\alpha}\iota_{\beta})\,\epsilon_{\dot\alpha\dot\beta}, \qquad
\omega^{3}_{ab}=2\i\, o_{(\alpha}\,\iota_{\beta)}\,\epsilon_{\dot\alpha\dot\beta}\,. 
\end{align}
These are all closed ($\d\omega^i=0$ for $i=1,2,3$) and ASD with respect to the flat metric $\eta_{ab}$ ($\star_{\eta}\omega^{i}_{ab}=-\omega^{i}_{ab}$), with the convention
\be\label{hodgestar}
\star_{\eta}\omega^i_{ab}=\frac{1}{2}\varepsilon_{ab}{}^{cd}\,\omega^{i}_{cd}\,, \qquad \varepsilon_{abcd}=\epsilon_{\alpha\gamma}\,\epsilon_{\beta\delta}\,\epsilon_{\dot\alpha\dot\delta}\,\epsilon_{\dot\beta\dot\gamma}-\epsilon_{\alpha\delta}\,\epsilon_{\beta\gamma}\,\epsilon_{\dot\alpha\dot\gamma}\,\epsilon_{\dot\beta\dot\delta}\,.
\ee
However, these 2-forms are \emph{not} ASD with respect to the curved metric $g_{ab}$ given by \eqref{SDMaxMet3}.

The correct hyperk\"ahler structure for $g_{ab}$ is contained in the following:
\begin{proposition}\label{prop:newHK}
Consider the complex hyperk\"ahler metric $g_{ab}$ defined in terms of a dual twistor quadric \eqref{quadric} by \eqref{SDMaxMet3}. Then:
\begin{enumerate}
\item\label{item:newHK} With $\omega^{i}_{ab}$ defined by \eqref{flatHK}, a hyperk\"ahler structure for $g_{ab}$ is given by 
\begin{align}\label{newHK}
\Omega^{1}_{ab}=\omega^{1}_{ab} + \tfrac{\i}{2}\,\varphi_{\dot\alpha\dot\beta}\,\epsilon_{\alpha\beta}\,, \qquad
\Omega^{2}_{ab}= \omega^{2}_{ab} - \tfrac{1}{2}\,\varphi_{\dot\alpha\dot\beta}\,\epsilon_{\alpha\beta}, \qquad
\Omega^{3}_{ab}= \omega^{3}_{ab}.
\end{align}
\item\label{item:3hol} The vector fields $\xi^{a}$ and $t^{a}$ defined by the quadric are Killing vectors for $g_{ab}$. Furthermore, $g_{ab}$ admits a tri-holomorphic Killing vector $\chi^{a}$, with $\chi^{a}=\xi^{a}$ if $a_{\alpha\beta}=0$ and $\chi^{a}=t^{a}$ if $a_{\alpha\beta}\neq0$.
\item\label{item:GH} The metric \eqref{SDMaxMet3} can be written in Gibbons-Hawking form 
\begin{align}\label{GH}
 g = V^{-1}\left(\d \psi+A\right)^{2}+V \left[ \d\mathcal{X}^2+\d\mathcal{Y}^2+\d\mathcal{Z}^2 \right]\,,
\end{align}
where $\partial_{\psi}=\chi^{a}\partial_{a}$ is the tri-holomorphic Killing vector from point \eqref{item:3hol}, and $\mathcal{X},\mathcal{Y},\mathcal{Z}$ are scalar fields defined by 
\begin{align}\label{deformedGHC0}
\d\mathcal{X} = -\chi\lrcorner\,\Omega^{1}\,, \qquad 
\d\mathcal{Y} = -\chi\lrcorner\,\Omega^{2}\,, \qquad 
\d\mathcal{Z} = -\chi\lrcorner\,\Omega^{3}\,.
\end{align}
\end{enumerate}
\end{proposition}

\proof We start with point (\ref{item:newHK}). Since $\omega^{i}_{ab}$ (defined by \eqref{flatHK}) are closed, and $F_{ab}=\varphi_{\dot\alpha\dot\beta} \epsilon_{\alpha\beta}$ is a SD Maxwell field and thus also closed (i.e., $\d F=0$ by the Bianchi identity), it follows that $\d\Omega^{i}=0$. So it remains to check whether $\star_{g}\Omega^{i}=-\Omega^{i}$, where $\star_{g}$ is the Hodge star operator with respect to $g_{ab}$. Since $g_{ab}$ differs from $\eta_{ab}$ by a Kerr-Schild term, its volume form is equal to that of the flat metric, $\varepsilon_{abcd}$ as given by \eqref{hodgestar}. Thus, we are left to compute
\be\label{HKstar1}
\star_{g}\Omega^{i}_{ab}=\frac{1}{2}\,\varepsilon_{abcd}\,g^{ce}\,g^{df}\,\Omega^{i}_{ef}\,,
\ee
where $g^{ab}=\eta^{ab}-h^{ab}$ for
\be\label{invKSmet}
h^{ab}=o^{\alpha}\,o^{\beta}\,\varphi^{\dot\alpha\dot\beta}\,,
\ee
the Kerr-Schild perturbation of the inverse metric.

The proof of (\ref{item:newHK}) now proceeds by direct calculation. Let us consider $\Omega^{1}_{ab}$ for example, with the $i=2,3$ cases following similarly. One can show that
\be\label{HKstar2}
\frac{1}{2}\varepsilon_{abcd}\,(\eta^{ce}-h^{ce})\,(\eta^{df}-h^{df})\,\omega^1_{ef}=-\omega^{1}_{ab}-\im\,\varphi_{\dot\alpha\dot\beta}\,\epsilon_{\alpha\beta}\,,
\ee
having used the explicit formulae of \eqref{flatHK} and the identity
\be\label{dyadident}
o_{\alpha}\,\iota_{\beta}-\iota_{\alpha}\,o_{\beta}=\epsilon_{\alpha\beta}\,.
\ee
Furthermore, the fact that $F_{ab}=\epsilon_{\alpha\beta}\varphi_{\dot\alpha\dot\beta}$ is null and SD gives
\be\label{HKstar3}
\frac{1}{2}\varepsilon_{abcd}\,(\eta^{ce}-h^{ce})\,(\eta^{df}-h^{df})F_{ef}=\frac{1}{2}\varepsilon_{abcd}\,\eta^{ce}\,\eta^{df}\,F_{ef}=F_{ab}\,,
\ee
which combines with \eqref{HKstar2} to show that $\star_{g}\Omega^{1}_{ab}=-\Omega^{1}_{ab}$. The two other cases follow by similar arguments.

Now consider (\ref{item:3hol}). The vector $\xi^{a}$ obeys $\cL_{\xi}g_{ab}=\cL_{\xi}h_{ab}$, for $h_{ab}= o_{\alpha}o_{\beta}\varphi_{\dot\alpha\dot\beta}$. Now, $\cL_{\xi}o_{\beta}=a_{\beta}{}^{\gamma} o_{\gamma}$ and $\cL_{\xi}\varphi_{\dot\alpha\dot\beta}=\xi^{c}\nabla_{c}\varphi_{\dot\alpha\dot\beta}$, so that
\be\label{triHol1}
\cL_{\xi}g_{ab} = o_{\alpha}\,o_{\beta}\,\xi^{c}\nabla_{c}\varphi_{\dot\alpha\dot\beta}+2\,\varphi_{\dot\alpha\dot\beta}\,a_{\alpha}{}^{\gamma}\,o_{\gamma}\,o_{\beta}\,.
\ee
A calculation using the formula \eqref{maxwellfield} for $\varphi_{\dot\alpha\dot\beta}$ reveals that 
\be\label{triHol2}
o_\alpha\,o_{\beta}\xi^{c}\nabla_{c}\varphi_{\dot\alpha\dot\beta} = -2a_{\alpha}{}^{\gamma}\, o_{\gamma}\,o_{\beta}\,\varphi_{\dot\alpha\dot\beta}\,,
\ee
from which it follows that $\cL_{\xi}g_{ab}=0$, so $\xi^a$ is a Killing vector for $g_{ab}$. Observe that if $a_{\alpha\beta}=0$, then $\cL_{\xi}o_{\alpha}=0$, so $\cL_{\xi}\Omega^{i}=0$ for $i=1,2,3$, and thus $\xi^{a}$ is tri-holomorphic. However, if $a_{\alpha\beta}\neq0$, then $\xi^{a}$ is \emph{not} tri-holomorphic. Furthermore, note that when $a_{\alpha\beta}=0$, it follows from \eqref{Killing2Tensor} that $t^{a}\propto\xi^a$, so that $t^a$ contains no new information.

For $t^{a}$, a simple calculation shows that $\cL_{t}o_{\beta}=0$, and a direct, more albeit tedious, calculation shows that $\cL_{t}\varphi_{\dot\alpha\dot\beta}=0$. Thus, $\cL_{t}h_{ab}=0$, so $\cL_{t}g_{ab}=0$, and $t^{a}$ is a Killing vector for $g_{ab}$. Moreover, it also follows that $\cL_{t}\Omega^{i}=0$, so $t^{a}$ is tri-holomorphic.

This establishes (\ref{item:3hol}), which means that $g_{ab}$ always possesses a tri-holomorphic Killing vector $\chi^a$. Any hyperk\"ahler metric with a tri-holomorphic Killing vector will necessarily admit a Gibbons-Hawking form \eqref{GH}, and the coordinates of the associated flat 3-space are the moment maps of the Killing field with respect to the hyperk\"ahler structure~\cite{Gibbons:1978tef}. In particular, this means that the flat coordinates on $\R^3$ are precisely given by \eqref{deformedGHC0}, establishing (\ref{item:GH}). \qed

\medskip

In practice, we will need expressions for the deformed Gibbons-Hawking coordinates \eqref{deformedGHC0} in terms of flat-space coordinates. These can be found as follows. The undeformed hyperk\"ahler structure \eqref{flatHK} defines the flat-space analogue of \eqref{deformedGHC0} via 
\begin{align}\label{undeformedGHC0}
\d{X}=-\chi\lrcorner\,\omega^{1}\,, \qquad 
\d{Y}=-\chi\lrcorner\,\omega^{2}\,, \qquad 
\d{Z}=-\chi\lrcorner\,\omega^{3}\,. 
\end{align}
Let $G$ be a scalar function solving 
\begin{align}\label{defG}
\nabla_{\alpha\dot\alpha}G=\varphi_{\dot\alpha\dot\beta}\,\chi_{\alpha}{}^{\dot\beta}\,. 
\end{align}
Then from \eqref{deformedGHC0}, \eqref{newHK} and \eqref{undeformedGHC0} one finds
\begin{align}\label{deformedGHC}
 \mathcal{X} = X - \i\,\frac{G}{2}\,, \qquad \mathcal{Y} = Y + \frac{G}{2}\,, \qquad \mathcal{Z} = Z\,,
\end{align}
for the deformed Gibbons-Hawking coordinates.

This has an immediately evident -- and important -- consequence when we demand that the curved metric $g_{ab}$ be real and Riemannian. In particular, \eqref{deformedGHC} implies that $g_{ab}$ must be defined on a \emph{new} real slice of the complexified spacetime; that is, a \emph{different} real slice to the one used for the flat metric. This is because $(X,Y)$ and $(\mathcal{X},\mathcal{Y})$ cannot be simultaneously real if $G\neq0$. Indeed, note that
\begin{align}
 \mathcal{X}+\i\mathcal{Y}=X+\i Y\,, \qquad \text{but} \qquad 
 \mathcal{X}-\i\mathcal{Y}=X-\i Y - \i G\,.
\end{align}
This fact means that the topology of the real manifold on which $g$ is defined can be different from that of $\R^4$.


\section{Dual twistor quadrics are self-dual black holes}
\label{sec:DTQClass}

At this point, we have established that every generic -- in the sense of \eqref{genericQuad} -- dual twistor quadric gives rise to a vacuum, self-dual (or hyperk\"ahler) metric which admits both single Kerr-Schild and Gibbons-Hawking forms. In this section, we classify all generic dual twistor quadrics leading to Riemannain hyperk\"ahler metrics, showing that there are three distinct classes. We then prove that these three classes correspond precisely to the three self-dual black hole metrics discussed in Section~\ref{sec:SDBH}.


\subsection{Classifying dual twistor quadrics}

We have seen in Section~\ref{sec: dual twistor quadrics to SD Kerr-Schild metrics} that any generic dual twistor quadric in $\mathbb{PT}^{*}$, with matrix representation
\begin{align}\label{quadric2}
 Q^{AB} = 
 \left( \begin{matrix} 
 c^{\dot\alpha\dot\beta} & b^{\dot\alpha}{}_{\beta} \\
 b_{\alpha}{}^{\dot\beta} & a_{\alpha\beta}
 \end{matrix} \right)\,,
\end{align}
gives rise to a hyperk\"ahler metric $g_{ab}$ \eqref{SDMaxMet3} via the SD null Maxwell field \eqref{maxwellfield}. This metric is also strictly conformally K\"ahler with respect to the orientation opposite to the hyperk\"ahler orientation. This conformal K\"ahler structure is encoded in the quadric through the Killing spinor $K^{\dot\alpha\dot\beta}$ (cf., \cite{Tod:1979tt,Pontecorvo:1992,Dunajski:2009dqa}).

When the parameter $\kappa\to0$ in the SD null Maxwell field, $g_{ab}$ passes smoothly to the flat metric $\eta_{ab}$, which nevertheless has a non-trivial conformal K\"ahler structure defined by $Q$ (cf., the conformal K\"ahler structure of SDPD \eqref{CKSPD}, which remains unmodified in the flat limit). If we require the resulting flat metric to be Euclidean-real, then $\PT^*$ inherits a real structure, and in order for the conformal K\"ahler structure to be real as well the quadric \emph{must} be preserved by the real structure of $\mathbb{PT}^{*}$. 

In terms of the components of $Q^{AB}$ in its matrix decomposition \eqref{quadric2}, this means that
\begin{align}\label{reality1}
\hat{a}_{\alpha\beta} = a_{\alpha\beta}\,, 
\qquad
\hat{b}_{\alpha}{}^{\dot\beta} = b_{\alpha}{}^{\dot\beta}\,,
\qquad
 \hat{c}^{\dot\alpha\dot\beta} = c^{\dot\alpha\dot\beta}\,.
\end{align}
Thus $a_{\alpha\beta}a^{\alpha\beta}\neq0$ and $c^{\dot\alpha\dot\beta}c_{\dot\alpha\dot\beta}\neq0$, which enables a decomposition into principal spinors as
\be\label{quadspindecomp1}
a_{\alpha\beta}=2\i\, a \, o_{(\alpha}\,\iota_{\beta)}\,, \qquad  c^{\dot\alpha\dot\beta}=2\i\, c \, o^{(\dot\alpha}\,\iota^{\dot\beta)}\,,
\ee
with $o_{\alpha}\iota^{\alpha}=1=o_{\dot\alpha}\iota^{\dot\alpha}$. The reality conditions \eqref{reality1} can then be stated as: $\hat{o}_{\alpha}=\iota_{\alpha}$, $\hat{o}_{\dot\alpha}=\hat{\iota}_{\dot\alpha}$ and $a,c\in\R$. 

Now, the choice of origin for (complexified) flat space $\M$ is arbitrary, so we are free to shift $x^{\alpha\dot\alpha}=\tilde{x}^{\alpha\dot\alpha}-x_{0}^{\alpha\dot\alpha}$, for $x_{0}^{\alpha\dot\alpha}$ an arbitrary constant 4-vector. Under this shift, the Killing spinor \eqref{Killingspinor} associated to the dual twistor quadric becomes
\be\label{Killingspinorshift}
K^{\dot\alpha\dot\beta} = a_{\alpha\beta}\,\tilde{x}^{\alpha\dot\alpha}\,\tilde{x}^{\beta\dot\beta}+2\left[2\, b_{\beta}{}^{(\dot\alpha}-a_{\alpha\beta}\,x_{0}^{\alpha(\dot\alpha}\right]\tilde{x}^{\dot\beta)\beta}  + c^{\dot\alpha\dot\beta}-2\, b_{\beta}{}^{(\dot\alpha}\,x_{0}^{\dot\beta)\beta} + a_{\alpha\beta}\,x_{0}^{\alpha\dot\alpha}\,x_{0}^{\beta\dot\beta}\,. 
\ee
Using this freedom, we see four distinct cases emerge, only three of which are interesting.

Firstly, suppose $a_{\alpha\beta}=0=b_{\alpha}{}^{\dot\beta}$. In this case the Killing spinor is constant for any choice of origin, $K^{\dot\alpha\dot\beta}=c^{\dot\alpha\dot\beta}$. This means that the dual twistor quadric is not generic in the sense of \eqref{genericQuad}, as $o^{\alpha}\frac{\partial Q}{\partial\omega^\alpha}=0$, so the construction of a SD Kerr-Schild metric via \eqref{twistorfunction} does not apply. Furthermore, as the Killing spinor is constant, the conformal K\"ahler structure associated to it is actually strictly K\"ahler; in light of \eqref{CKSPD}, it is clear that this cannot be associated to a SD black hole metric. This underlines the importance of the genericity condition \eqref{genericQuad}, and we discard this degenerate case from now on.

Next, suppose that $a_{\alpha\beta}=0$ while $b_{\alpha}{}^{\dot\alpha}\neq 0$. Then by appropriately choosing $x_{0}^{\alpha\dot\alpha}$ in \eqref{Killingspinorshift}, it is possible to remove the zeroth-order (i.e., constant) part of the Killing spinor altogether. This case is therefore captured by dual twistor quadrics of the form
\be\label{CaseA}
\mbox{Case A:} \qquad Q^{AB}=\left( \begin{matrix} 
 0 & b^{\dot\alpha}{}_{\beta} \\
 b_{\alpha}{}^{\dot\beta} & 0
 \end{matrix} \right)\,.
 \ee
This case will be analysed in section \ref{sec:SDTN} below.

If $a_{\alpha\beta}\neq0$, then two distinct cases emerge. By appropriately choosing $x_{0}^{\alpha\dot\alpha}$, the linear term in \eqref{Killingspinorshift} can be eliminated.
The two cases then correspond to whether it is possible to eliminate the zeroth order piece of the Killing vector simultaneously. In other words, the two cases are described by dual twistor quadrics of the form
\be\label{CaseB}
\mbox{Case B:} \qquad Q^{AB}=\left( \begin{matrix} 
 0 & 0 \\
 0 & a_{\alpha\beta}
 \end{matrix} \right)\,,
\ee
which will be analysed in section~\ref{sec:EH}, and
\be\label{CaseC}
\mbox{Case C:} \qquad Q^{AB}=\left( \begin{matrix} 
 c^{\dot\alpha\dot\beta} & 0 \\
 0 & a_{\alpha\beta}
 \end{matrix} \right)\,,
\ee
which will be analyzed in Section~\ref{sec:SDPD}.

Cases A, B and C form an exhaustive classification of all generic dual twistor quadrics, in the sense that all possible generic quadrics fall into precisely one of the three cases. We will now establish that each of these cases corresponds to one of the self-dual black hole metrics.


\subsection{Case A: Self-dual Taub-NUT}
\label{sec:SDTN}

Let us begin with dual twistor quadrics of the type Case A, described by \eqref{CaseA}. In this case, the Killing spinor is simply
\be\label{CaseAKS}
K^{\dot\alpha\dot\beta}=2\,b_{\beta}{}^{(\dot\alpha}\,x^{|\beta|\dot\beta)}\,,
\ee
and the reality condition \eqref{reality1} ensures that we can write
\be\label{CaseAb}
b_{\alpha}{}^{\dot\beta}=\frac{1}{2\,\sqrt{2}}\left(o_{\alpha}\,o^{\dot\beta}+\iota_{\alpha}\,\iota^{\dot\beta}\right)\,.
\ee
A straightforward calculation then gives 
\be\label{CaseAK^2}
K^{\dot\alpha\dot\beta}\, K_{\dot\alpha\dot\beta}=2\left(x^2+y^2+z^2\right)=2\,r^2\,,
\ee
while the roots defined by \eqref{roots} are seen to be
\be\label{CaseAroots}
\zeta_{\pm}=\frac{z\mp r}{x+\im y}\,.
\ee
From this, the principal spinors \eqref{prinspin2}
\be\label{CaseAPS}
\alpha^{\dot\alpha}=\sqrt{\frac{y-\im x}{2}}\left(o^{\dot\alpha}+\zeta_{-}\,\iota^{\dot\alpha}\right)\,, \qquad \beta^{\dot\alpha}=\sqrt{\frac{y-\im x}{2}}\left(o^{\dot\alpha}+\zeta_{+}\,\iota^{\dot\alpha}\right)\,,
\ee
of $K^{\dot\alpha\dot\beta}$ are easily determined. 

Making use of the formula \eqref{maxwellfield2} we can now easily obtain an expression for the null SD Maxwell field associate to dual twistor quadrics covered by Case A:
\be\label{CaseAMax}
\varphi_{\dot\alpha\dot\beta}=\frac{\kappa}{r}\,(o_{\dot\alpha}+\zeta_{-}\,\iota_{\dot\alpha}) \,(o_{\dot\beta}+\zeta_{-}\,\iota_{\dot\beta})\,.
\ee
This in turn in allows us to read off the associated SD Kerr-Schild metric; in the flat coordinate system
\be\label{flatcoordconv}
x^{\alpha\dot\alpha}=\left(\begin{array}{cc}
u & w \\
\tilde{w} & v
\end{array}\right)=\frac{1}{\sqrt{2}}\left(\begin{array}{cc}
\tau-\im z & \im x-y \\
\im x+y & \tau-\im z
\end{array}\right)\,,
\ee
the non-trivial part of this metric is given by
\be\label{CaseA-KS}
h^{(A)}_{ab}\,\d x^{a}\,\d x^{b}=\frac{\kappa}{r}\left(\d u+\zeta_{-}\,\d w\right)^2\,.
\ee
In other words, every dual twistor quadric of type A leads to a hyperk\"ahler, Kerr-Schild metric of the form \eqref{CaseA-KS}.

We now observe the following:
\begin{proposition}\label{prop:TN}
The hyperk\"ahler metric defined by dual twistor quadrics of type A \eqref{CaseA} is isometric to the self-dual Taub-NUT metric.
\end{proposition}

\proof By Proposition~\ref{prop:newHK}, the hyperk\"ahler Kerr-Schild metric associated to any dual twistor quadric can be written in Gibbons-Hawking form \eqref{GH}. By the same proposition, for Case A (where $a_{\alpha\beta}=0$) the tri-holomorphic Killing vector of this Gibbons-Hawking metric will be given by
\be\label{CaseAchi}
\chi^{\alpha\dot\alpha}\equiv\xi^{\alpha\dot\alpha}=\frac{1}{\sqrt{2}}\left(o_{\alpha}\,o^{\dot\beta}+\iota_{\alpha}\,\iota^{\dot\beta}\right)\,.
\ee
To determine the Gibbons-Hawking form of the metric, we then need to find the deformed coordinates $(\mathcal{X},\mathcal{Y},\mathcal{Z})$, and an expression for the potential $V$ in terms of them. 

These deformed coordinates are determined by the `undeformed' coordinates $(X,Y,Z)$ defined by \eqref{undeformedGHC0} of the flat hyperk\"ahler structure \eqref{flatHK} along with a scalar function $G$ solving \eqref{defG}, via \eqref{deformedGHC}. Using \eqref{undeformedGHC0} and \eqref{flatHK} with $\chi^{a}=\xi^{a}$ gives
\begin{align}\label{undefGHCEH}
X = x\,, \qquad Y= y\,, \qquad Z=z\,,
\end{align}
so the undeformed coordinates can be identified with the standard Cartesian coordinates on $\R^3$.

To determine $G$, one uses \eqref{CaseAMax} in \eqref{defG} to obtain the differential equation
\be\label{CaseAdG}
\d G=\frac{\kappa}{4\, r}\left[2\i\, \zeta_{-}\,\d{z}+\i\left(1-\zeta_{-}^2\right)\d{x}+\left(1+\zeta_{-}^2\right)\d{y}\right]\,,
\ee
which can be integrated to
\begin{align}\label{CaseAG}
G = \i\,\kappa\,\frac{z+r}{x+\i y}\,,
\end{align}
using \eqref{CaseAroots}. The deformed Gibbons-Hawking coordinates are then read off from \eqref{deformedGHC}: 
\begin{align}\label{deformedGHC-SDTN}
 \mathcal{X} = x + \frac{\kappa}{2}\left(\frac{z+r}{x+\i y} \right), \qquad 
 \mathcal{Y} = y + \frac{\i\,\kappa}{2}\left(\frac{z+r}{x+\i y} \right), \qquad 
 \mathcal{Z} = z\,, 
\end{align}
and it remains to determine the potential $V$. 

Using the facts that $V^{-1}=g_{ab}\xi^a\xi^b$, $\eta_{ab}\xi^a\xi^b=1$ and $o_{\alpha}\alpha_{\dot\alpha}\xi^{\alpha\dot\alpha}=2^{-1/2}$, it follows that 
\begin{align}\label{VTN0}
 V^{-1} = 1+\frac{\kappa}{2\,r}\,.
\end{align}
To express this in the deformed coordinates \eqref{deformedGHC-SDTN}, first define
\begin{align}
 \mathcal{R}:=\sqrt{\mathcal{X}^2+\mathcal{Y}^2+(\mathcal{Z}-\tfrac{\kappa}{2})^2} \; .
\end{align}
Then a short calculation gives $\mathcal{R}=r+\frac{\kappa}{2}$, and \eqref{VTN0} becomes
\begin{align}\label{VTN}
 V(\mathcal{X},\mathcal{Y},\mathcal{Z}) = 1 - \frac{\kappa}{2\,\mathcal{R}}\,.
\end{align}
This establishes that dual twistor quadrics of type A give rise to metrics isometric to a Gibbons-Hawking metric \eqref{GH} with $V$ given by \eqref{VTN}. This is an ALF, single-centred solution, so it is necessarily the SDTN metric with mass $M=-\kappa/4$ by comparison with \eqref{SDTN-GH}. Furthermore, imposing regularity requires $M>0$, which in turn means that $\kappa<0$. \qed

\medskip

This result then implies the easy
\begin{corollary}
The SDTN metric is isometric to the Kerr-Schild metric defined by \eqref{CaseA-KS}
\end{corollary}
This Kerr-Schild form of SDTN agrees with that found recently in~\cite{Kim:2024dxo}.


\subsection{Case B: Eguchi-Hanson}
\label{sec:EH}

Now consider dual twistor quadrics falling into Case B, described by \eqref{CaseB}. In this case, the Killing spinor is 
\be\label{CaseBKS}
K^{\dot\alpha\dot\beta}=a_{\alpha\beta}\,x^{\alpha\dot\alpha}\,x^{\beta\dot\beta}=2\im\,a\,o_{(\alpha}\,\iota_{\beta)}\,x^{\alpha\dot\alpha}\,x^{\beta\dot\beta}\,,
\ee
using the decomposition \eqref{quadspindecomp1}. From this, one can immediately identify the principal spinors
\be\label{CaseBPS}
\alpha^{\dot\alpha}=\sqrt{2\im\,a}\,x^{\alpha\dot\alpha}\,\iota_{\alpha}\,, \qquad \beta^{\dot\alpha}=\sqrt{2\im\,a}\,x^{\alpha\dot\alpha}\,o_{\alpha}\,,
\ee
and observe that
\be\label{CaseBPS2}
\alpha^{\dot\alpha}=-\sqrt{2\im\,a}\,v\left(o^{\dot\alpha}+\zeta_{-}\,\iota^{\dot\alpha}\right)\,, \qquad \mbox{for }\: \zeta_{-}:=-\frac{\tilde{w}}{v}\,,
\ee
in the flat coordinates \eqref{flatcoordconv}. It is then straightforward to evaluate the associated null SD Maxwell field
\be\label{CaseBMax}
\varphi_{\dot\alpha\dot\beta}=-\frac{\kappa\,v^2}{a^3\,(x^c\,x_c)^3}(o_{\dot\alpha}+\zeta_{-}\,\iota_{\dot\alpha})\,(o_{\dot\beta}+\zeta_{-}\,\iota_{\dot\beta})\,.
\ee
This in turn indicates that dual twistor quadrics of type B lead to SD Kerr-Schild metrics with
\be\label{CaseB-KS}
h^{(B)}_{ab}\,\d x^a\,\d x^b=-\frac{\kappa\,v^2}{a^3\,(x^c\,x_c)^3}\,\left(\d u+\zeta_-\,\d w\right)^2\,,
\ee
being the non-trivial part of the metric.

\begin{proposition}\label{prop:EH}
The hyperk\"ahler metric defined by dual twistor quadrics of type B \eqref{CaseB} is isometric to the Eguchi-Hanson metric.
\end{proposition}

\proof 
As in Case A, we proceed by converting the metric to Gibbons-Hawking form using Proposition~\ref{prop:newHK}. 
In Case B, this Proposition implies that, since $a_{\alpha\beta}\neq0$, the tri-holomorphic Killing vector associated with the hyperk\"ahler metric is 
\begin{align}
\chi^{a}\equiv t^a=H^{ab}\,\xi_{b}\,.
\end{align}
It is convenient to compute this explicitly in the more general Case C where $a_{\alpha\beta}\neq0$ and $c_{\dot\alpha\dot\beta}\neq0$ (but $b_{\alpha\dot\alpha}=0$): a calculation gives
\begin{align}\label{tanotzero}
 t^{\alpha\dot\alpha} = -2\,a^2\, c^{\dot\alpha}{}_{\dot\beta}\,x^{\alpha\dot\beta}\,.
\end{align}
Since the quadric in Case B has $c_{\dot\alpha\dot\beta}=0$, it appears that, in principle, $t^{\alpha\dot\alpha}$ vanishes for this case. 

To avoid this issue, note that we can rescale the vector \eqref{tanotzero} by any constant and it continues to be a Killing vector. In particular, using \eqref{quadspindecomp1} $c^{\dot\alpha}{}_{\dot\beta}=\i\, c\,(o^{\dot\alpha}\iota_{\dot\beta}+\iota^{\dot\alpha}o_{\dot\beta})$, we can rescale $t^a$ by $(-2a^2c)^{-1}$ to obtain the tri-holomorphic Killing vector
\begin{align}\label{tEH}
 \chi^{\alpha\dot\alpha} =  \i\, (o^{\dot\alpha}\iota_{\dot\beta}+\iota^{\dot\alpha}o_{\dot\beta})\,x^{\alpha\dot\beta}\,,
\end{align}
which is non-vanishing for Case B, as desired.

We now need to find the deformed coordinates $(\mathcal{X},\mathcal{Y},\mathcal{Z})$, and the expression for the potential $V$ in terms of them. The former are given by \eqref{deformedGHC}, so we need to find $(X,Y,Z)$ and the function $G$. Using \eqref{undeformedGHC0} and \eqref{flatHK}, and the expression \eqref{tEH} for $\chi^a$, one finds the undeformed coordinates: 
\begin{align}\label{undefGHCEH0}
X = uw-v\tilde{w}\,, \qquad Y=-\i\, (uw+v\tilde{w})\,, \qquad Z=uv+w\tilde{w}\,.
\end{align}
To compute the function $G$, one uses \eqref{defG}, \eqref{CaseBMax} and \eqref{tEH} to find the differential equation
\begin{align}
\d G = -\i\kappa\,\frac{v(uv+w\tilde{w})\,\d\tilde{w}-2v^2\tilde{w}\,\d{u}+2v\tilde{w}^2\,\d{w}-\tilde{w}(uv+w\tilde{w})\,\d{v}}{8\,a^3\,(uv-w\tilde{w})^3}\,.
\end{align}
This can be integrated to give
\begin{align}
 G = -\frac{\i\,\kappa}{8\,a^3}\,\frac{v\,\tilde{w}}{(uv-w\tilde{w})^2} 
 = \frac{\i\, \kappa}{16\,a^3}\frac{(X-\i Y)}{(X^2+Y^2+Z^2)}\,,
\end{align}
having used \eqref{undefGHCEH0} to obtain the second equality. 

Defining $R=\sqrt{X^2+Y^2+Z^2}$ and using \eqref{deformedGHC}, the deformed Gibbons-Hawking coordinates are then
\begin{align}\label{defGHCEH}
\mathcal{X} = X + \frac{\kappa}{32a^3}\,\frac{(X-\i Y)}{R^2}\,, \qquad
\mathcal{Y} = Y + \frac{\i\kappa}{32a^3}\,\frac{(X-\i Y)}{R^2}\,, \qquad 
\mathcal{Z} = Z\,.
\end{align}
The associated potential is then obtained from 
\[
V^{-1} = g_{ab}\,\chi^{a}\,\chi^{b}=
2(uv-w\tilde{w})+\frac{\kappa}{8\,a^3}\,\frac{(uv+w\tilde{w})^2}{(uv-w\tilde{w})^3}\,.
\]
Using \eqref{undefGHCEH0}, this gives 
\[
V =\frac{R^3}{2(R^4+\frac{\kappa}{16a^3}\,Z^2)}\,. 
\]
Now, define $Z_0^2:=-\kappa/(16a^3)$ and
\begin{align}\label{RpmEH}
\mathcal{R}_{\pm}:=\sqrt{\mathcal{X}^2+\mathcal{Y}^2+(\mathcal{Z} \pm Z_0)^2}\,.
\end{align}
Using \eqref{defGHCEH}, a short calculation gives $\mathcal{R}_{\pm}=R\pm Z_0 Z/R$, so $\mathcal{R}_{+}+\mathcal{R}_{-}=2R$ and $\mathcal{R}_{+}\mathcal{R}_{-}=(R^4-Z_0^2Z^2)/R^2$. This means that the expression for $V$ can be re-written as
\begin{align}\label{VEH1}
 V(\mathcal{X},\mathcal{Y},\mathcal{Z}) = \frac{1}{4}\left(\frac{1}{\mathcal{R}_{+}} + \frac{1}{\mathcal{R}_{-}}\right).
\end{align}
The metric is then \eqref{GH} with the potential $V$ given by \eqref{VEH1}; that is, an ALE two-centred Gibbons-Hawking metric with equal masses. 
By comparison with \eqref{GHEH} -- \eqref{EHparam}, it must be the Eguchi-Hanson metric. \qed

\medskip

Having identified the hyperk\"ahler metrics of Case B with Eguchi-Hanson, we can now \emph{a posteriori} recognize the Kerr-Schild metric \eqref{CaseB-KS} as the known Kerr-Schild form of the Eguchi-Hanson metric~\cite{Sparling:1981nk,Tod:1982mmp}.


\subsection{Case C: Self-dual Pleba\'nski-Demia\'nski}
\label{sec:SDPD}

Finally, consider the dual twistor quadrics covered by Case C, described by \eqref{CaseC}. This case is, by far, the most technically complicated to analyze. Using \eqref{quadspindecomp1}, the Killing spinor associated to Case C is
\be\label{CaseCKS}
K^{\dot\alpha\dot\beta}=2\im\left(a\,o_{(\alpha}\,\iota_{\beta)}\,x^{\alpha\dot\alpha}\,x^{\beta\dot\beta}+c\,o^{(\dot\alpha}\,\iota^{\dot\beta)}\right)\,,
\ee
from which it follows that

\be\label{CaseCK^2}
K^{\dot\alpha\dot\beta}\,K_{\dot\alpha\dot\beta}=
2\left[a^2\,(uv-w\tilde{w})^2-2\,a\,c\,(uv+w\tilde{w})+c^2 \right].
\ee
The principal spinors of $K^{\dot\alpha\dot\beta}$ are found to be
\begin{align}
\alpha^{\dot\alpha} = \sqrt{2\i\, a\, wv}\,(o^{\dot\alpha}+\zeta_{-}\,\iota^{\dot\alpha})\,, 
\qquad 
\beta^{\dot\alpha} = \sqrt{2\i\, a\, wv}\,(o^{\dot\alpha}+\zeta_{+}\,\iota^{\dot\alpha})\,,
\end{align}
where the roots $\zeta_{\pm}$ are 
\begin{align}\label{rootsPD}
\zeta_{\pm} = \frac{1}{2awv}\left[c-a( uv+w\tilde{w}) \pm \sqrt{a^2(uv-w\tilde{w})^2-2ac(uv+w\tilde{w})+c^2}  \right]\,,
\end{align}
as defined by \eqref{roots}.

To compute the associated SD null Maxwell field, note that 
\be\label{CaseCxi}
\xi_{\alpha\dot\alpha}=4\im\,a\,o_{(\alpha}\,\iota_{\beta)}\,x^{\beta}{}_{\dot\alpha}\,,
\ee
from which it follows, by \eqref{maxwellfield2}, that 
\begin{align}\label{CaseCMax}
\varphi_{\dot\alpha\dot\beta} =
\frac{-\kappa\, (o_{\dot\alpha}+\zeta_{-}\iota_{\dot\alpha})\,(o_{\dot\beta}+\zeta_{-}\iota_{\dot\beta})}{8 a^2\, (u+\zeta_{-}w)^2\,\sqrt{a^2(uv-w\tilde{w})^2-2ac\,(uv+w\tilde{w})+c^2}} \, .
\end{align}
Thus, the Kerr-Schild perturbation associated to dual twistor quadrics of type C is 
\begin{align}\label{CaseC-KS}
h^{(C)}_{ab}\,\d{x}^{a}\,\d{x}^{b} ={}& 
\frac{-\kappa\,(\d{u}+\zeta_{-}\d{w})^2 }{8 a^2\, (u+\zeta_{-}w)^2\,\sqrt{a^2(uv-w\tilde{w})^2-2ac\,(uv+w\tilde{w})+c^2}} \,.
\end{align}

\begin{proposition}
The hyperk\"ahler metric defined by dual twistor quadrics of type C \eqref{CaseC} is isometric to the self-dual Pleba\'nski-Demia\'nski metric.
\end{proposition}

\proof As in previous cases, we first convert the metric to Gibbons-Hawking form using Proposition~\ref{prop:newHK}. In Case C, the tri-holomorphic Killing vector associated with the hyperk\"ahler metric is \eqref{tanotzero}, or equivalently \eqref{tEH} after dividing by a factor of $(-2ac^2)$. To streamline calculations, we take the simpler option, so that the tri-holomorphic Killing vector $\chi^{\alpha\dot\alpha}$ is given by \eqref{tEH}. 

To find the deformed Gibbons-Hawking coordinates $(\mathcal{X},\mathcal{Y},\mathcal{Z})$ and the expression of $V$ in terms of them, we first need $(X,Y,Z)$ and the function $G$. Since the Killing vector is \eqref{tEH}, the un-deformed coordinates $(X,Y,Z)$ are the same as in the Eguchi-Hanson case: 
\begin{align}\label{undefGHCPD}
X = uw-v\tilde{w}\,, \qquad Y=-\i (uw+v\tilde{w})\,, \qquad Z=uv+w\tilde{w}\,.
\end{align}
To compute $G$, we use \eqref{defG}, \eqref{CaseCMax} and \eqref{tEH}; a short calculation gives the differential equation
\begin{align}
\d G = -\i\kappa\,\frac{(\zeta_{-}v-\tilde{w})\,\d{u} +(u-\zeta_{-}w)\,\d\tilde{w}+\zeta_{-}(\zeta_{-}v-\tilde{w})\,\d{w} + \zeta_{-}(u-\zeta_{-}w)\,\d{v} }{8a^2\,(u+\zeta_{-}w)^2\,\sqrt{a^2\,(uv-w\tilde{w})^2-2ac\,(uv+w\tilde{w})+c^2}}\,.
\end{align}
Despite the apparent complexity of this equation, it has a remarkably simple solution:
\begin{align}\label{GSDPD}
G = \frac{\i\,\kappa}{8a^2c}\frac{(\tilde{w}+\zeta_{-}\,v)}{(u+\zeta_{-}\,w)}\,,
\end{align}
as can be checked by direct calculation.

To express $G$ in terms of the coordinates $(X,Y,Z)$ given by \eqref{undefGHCPD}, define 
\begin{align}\label{RPD}
 R:=\sqrt{X^2+Y^2+Z^2} = uv-w\tilde{w}\,,
\end{align}
where the second equality uses the explicit expression \eqref{undefGHCPD}. Then the roots \eqref{rootsPD} become
\be\label{rootsPD2}
\begin{split}
\zeta_{\pm} =& \frac{-1}{2vw}\left[Z-\tfrac{c}{a} \pm \sqrt{R^2-\tfrac{2c}{a}Z+(\tfrac{c}{a})^2} \,  \right] \\
={}& \frac{-1}{2vw}\left[Z-\tfrac{c}{a} \pm \sqrt{X^2+Y^2+(Z-\tfrac{c}{a})^2} \,  \right].
\end{split}
\ee
From this, it follows that:
\begin{align}
\frac{\tilde{w}+\zeta_{-}v}{u+\zeta_{-}w} = \frac{v}{w}\,\frac{\left[2w\tilde{w}-(Z-\frac{c}{a}-\sqrt{X^2+Y^2+(Z-\tfrac{c}{a})^2} \, ) \right]}{\left[2uv-(Z-\frac{c}{a}-\sqrt{X^2+Y^2+(Z-\tfrac{c}{a})^2} \, ) \right]}\,.
\end{align}
Now, using \eqref{undefGHCPD} and \eqref{RPD}, it is straightforward to deduce the identities 
\begin{align}
2w\tilde{w}-Z = -R\,, \qquad 2uv-Z=R\,, \qquad \frac{Z+R}{X+\i Y} = \frac{v}{w}\,,
\end{align}
which lead to
\begin{align}\label{PDident0}
\frac{\tilde{w}+\zeta_{-}v}{u+\zeta_{-}w} = 
\frac{(Z+R)}{(X+\i Y)}\frac{\left[-R+\frac{c}{a}+\sqrt{X^2+Y^2+(Z-\tfrac{c}{a})^2} \right]}{\left[R+\frac{c}{a}+\sqrt{X^2+Y^2+(Z-\tfrac{c}{a})^2} \right]}\,.
\end{align}
Finally, using another identity
\begin{multline}
(Z+R)\frac{\left[-R+\frac{c}{a}+\sqrt{X^2+Y^2+(Z-\tfrac{c}{a})^2} \right]}{\left[R+\frac{c}{a}+\sqrt{X^2+Y^2+(Z-\tfrac{c}{a})^2} \right]} 
\\
= Z-\frac{a}{c}R^2+\frac{a}{c}R\sqrt{X^2+Y^2+(Z-\tfrac{c}{a})^2}\,,
\end{multline}
in conjunction with \eqref{PDident0} in \eqref{GSDPD}, gives the expression:
\begin{align}
G = \frac{-\i\,\kappa}{8ac^2}\left[\frac{X^2+Y^2+Z\,(Z-\frac{c}{a})-\sqrt{X^2+Y^2+Z^2}\sqrt{X^2+Y^2+(Z-\frac{c}{a})^2}}{(X+\i Y)}\right]\,,
\end{align}
for $G$ in terms of the un-deformed coordinates. Using \eqref{deformedGHC} then immediately gives 
\begin{equation}\label{defGHC-PD}
\begin{aligned}
\mathcal{X} ={}& X - \frac{\kappa}{16ac^2}\left[\frac{X^2+Y^2+Z(Z-\frac{c}{a})-\sqrt{X^2+Y^2+Z^2}\sqrt{X^2+Y^2+(Z-\frac{c}{a})^2}}{(X+\i Y)}\right], \\
\mathcal{Y} ={}& Y - \frac{\i\kappa}{16ac^2}\left[\frac{X^2+Y^2+Z(Z-\frac{c}{a})-\sqrt{X^2+Y^2+Z^2}\sqrt{X^2+Y^2+(Z-\frac{c}{a})^2}}{(X+\i Y)}\right], \\
\mathcal{Z} ={}& Z\,,
\end{aligned}
\end{equation}
for the deformed Gibbons-Hawking coordinates.

To obtain the hyperk\"ahler metric in these Gibbons-Hawking coordinates, we next need to compute the potential $V=(g_{ab}\chi^a\chi^b)^{-1}$ and express it in terms of \eqref{defGHC-PD}. 
Recalling that the curved metric is $g_{ab}=\eta_{ab}+o_{\alpha}o_{\beta}\varphi_{\dot\alpha\dot\beta}$, it follows that
\begin{align}
\nonumber V^{-1} ={}& \eta_{ab}\,\chi^{a}\,\chi^{b}+o_{\alpha}o_{\beta}\,\varphi_{\dot\alpha\dot\beta}\,\chi^{\alpha\dot\alpha}\,\chi^{\beta\dot\beta} \\
={}& x_{a}x^{a}+ \frac{\kappa}{\sqrt{2\,K^{\dot\gamma\dot\delta}\,K_{\dot\gamma\dot\delta}}}\,\frac{(k_{a}\chi^{a})^2}{(k_{b}\xi^{b})^2}\,,
\label{VPDcalc0}
\end{align}
where in the second line we used that $\eta_{ab}\chi^{a}\chi^{b}=x_{a}x^{a}$, together with the identity \eqref{maxwellfield2} and the definition $k_{a}=o_{\alpha}\alpha_{\dot\alpha}$. 

The calculation is now simplified by noticing that $k^{a}$ is an eigenvector of the Killing tensor $H_{ab}$: recalling the expression \eqref{Killing2Tensor}, one has that
\begin{align}
H_{ab}\,k^{b} ={}& a_{\alpha\beta}\,K_{\dot\alpha\dot\beta}\,o^{\beta}\,\alpha^{\dot\beta}-\frac{\xi^{c}\xi_{c}}{8}\,k_{a} \\
={}& \left( \frac{\i\, a}{2}\,\alpha_{\dot\alpha}\,\beta^{\dot\alpha}-\frac{\xi^{c}\xi_{c}}{8} \right) k_{a},
\end{align}
where in the second line we used that $a_{\alpha\beta}o^{\beta}=-\i ao_{\alpha}$ and $K_{\dot\alpha\dot\beta}\alpha^{\dot\beta}=-\frac{1}{2}\alpha_{\dot\beta}\beta^{\dot\beta}\alpha_{\dot\alpha}$. Since $K^{\dot\alpha\dot\beta}=\alpha^{(\dot\alpha}\beta^{\dot\beta)}$, it follows that $K^{\dot\alpha\dot\beta}K_{\dot\alpha\dot\beta}=-\frac{1}{2}(\alpha_{\dot\alpha}\beta^{\dot\alpha})^{2}$, so $\sqrt{2K^{\dot\alpha\dot\beta}K_{\dot\alpha\dot\beta}}=\i \alpha_{\dot\alpha}\beta^{\dot\alpha}$. Defining 
\begin{align}\label{PPD}
P:=\sqrt{X^2+Y^2+(Z-\tfrac{c}{a})^2},
\end{align}
it follows that $\i \alpha_{\dot\alpha}\beta^{\dot\alpha}=\sqrt{2K^{\dot\alpha\dot\beta}K_{\dot\alpha\dot\beta}}=2aP$. 

Using $\xi^c\xi_c=4a^2x^cx_c=8a^2R$ -- where $R$ was defined in \eqref{RPD} -- then gives
\begin{align}
H_{ab}\,k^{b} = a^2\,(P-R)\,k_{a}\,.
\end{align}
Therefore, since $\chi_{a}=\frac{-1}{2a^2c}t_{a}=\frac{-1}{2a^2c}H_{ab}\xi^{b}$, one has
\begin{align*}
k_{a}\,\chi^{a} = \frac{-1}{2a^2c}\,H_{ab}\,\xi^{b}\,k^{a}=\frac{-1}{2c}\,(P-R)\,k_{a}\,\xi^{a}\,.
\end{align*}
Feeding this into \eqref{VPDcalc0}, and again using that $\sqrt{2K^{\dot\alpha\dot\beta}K_{\dot\alpha\dot\beta}}=2aP$, gives
\begin{align}
V^{-1} = 2R + \frac{\kappa}{8ac^2}\,\frac{(P-R)^2}{P}\,,
\end{align}
for the inverse of the scalar potential.

To proceed further, it is convenient to define the parameter 
\begin{align}
\varepsilon:=\sqrt{c^2-\frac{\kappa}{4a}}\,, 
\end{align}
and assume for the moment that $c^2\neq\frac{\kappa}{4a}$ -- that is, that $\varepsilon\neq0$. Letting
\begin{align}\label{RpmPD}
\mathcal{R}_{\pm}:=\sqrt{\mathcal{X}^2+\mathcal{Y}^2+(\mathcal{Z}-\tfrac{(c\mp\varepsilon)}{2a})^2}\,,
\end{align}
one can check that 
\begin{align}
\mathcal{R}_{\pm}=\frac{1}{2}\left[ (P+R) \mp \frac{\varepsilon}{c}\,(P-R) \right]\,.
\end{align}
Using $\kappa=4a(c^2-\varepsilon^2)$, we now compute:
\begin{align}
\nonumber V ={}& \frac{8ac^2\, P}{16ac^2\, PR + \kappa\, (P-R)^2} \\
\nonumber ={}& \frac{8c^2\, P}{4ac^2\, [(P+R)^2-(P-R)^2]+4a\,(c^2-\varepsilon^2)\,(P-R)^2 } \\
\nonumber ={}& \frac{2c^2\, P}{c^2\, (P+R)^2 -\varepsilon^2\,(P-R)^2 } \\
\nonumber ={}& \frac{2c^2\, P}{[c\,(P+R) +\varepsilon\,(P-R)]\,[c\,(P+R) - \varepsilon\,(P-R)] } \\
\nonumber ={}& \frac{c}{2\varepsilon}\left(\frac{(\varepsilon-c)\,[c\,(P+R) - \varepsilon\,(P-R)]+(\varepsilon+c)\,[c\,(P+R) + \varepsilon\,(P-R)]\}}{[c\,(P+R) +\varepsilon\,(P-R)]\,[c\,(P+R) - \varepsilon\,(P-R)] }\right) \\
={}& \frac{\mu_{+}}{\mathcal{R}_{+}}+\frac{\mu_{-}}{\mathcal{R}_{-}}
\end{align}
where
\begin{align}\label{massesSDPD2}
\mu_{\pm}:=\frac{\varepsilon\pm c}{4\varepsilon}.
\end{align}
Since it was assumed $\varepsilon\neq0$, the potential then corresponds to an ALE two-centred Gibbons-Hawking metric with different masses. In view of the construction in section \ref{sec:SDPD0}, it must then be isometric to the non-degenerate case of the self-dual Pleba\'nski-Demia\'nski metric. Note that, as required for the two masses \eqref{masseps} of SDPD, $\mu_{\pm}$ are not independent, but rather are both controlled by $\varepsilon$ and $c$.

Finally, consider the degenerate limit $\varepsilon\to0$ in the above formulae: in terms of the dual twistor quadric, this means that $c^2=\frac{\kappa}{4a}$. From \eqref{RpmPD} and \eqref{massesSDPD2}, it follows that in this limit the centres coincide ($\mathcal{R}_{+}\to\mathcal{R}_{-}$) while the masses diverge. This is precisely the behaviour of the degenerate case of SDPD studied in Section~\ref{sec:SDPD0}, as required. \qed

\medskip

A surprising consequence of this result is the following:
\begin{corollary}
The SDPD metric is isometric to the Kerr-Schild metric defined by \eqref{CaseC-KS}.
\end{corollary}
To our knowledge, the fact that the SDPD metric admits a single Kerr-Schild form was not known previously. Indeed, this seems to be quite remarkable, in light of how complicated the metric is when written in Pleba\'nski-Demia\'nski,  SU$(\infty)$ Toda or even Gibbons-Hawking coordinates.


\section{Discussion}
\label{sec:Discussion}

In this paper, we proved that all self-dual black holes, together with their geometric properties, are entirely encoded in data from {\em flat} spacetime, by developing a construction that depends purely on the {\em dual} twistor space of Euclidean 4-space. In more technical terms, we showed that given a holomorphic quadratic variety in (an open subset of) complex projective 3-space (i.e., a dual twistor quadric), one can explicitly construct a hyperk\"ahler metric which is also strictly conformally K\"ahler with the {\em opposite} orientation. This procedure also generates all of the symmetries and `hidden' symmetries of the solutions (Killing spinors, Killing tensors and toric structures). Furthermore, we classified all such dual twistor quadrics, showing that only three non-trivial cases can arise, corresponding to the self-dual Taub-NUT, Eguchi-Hanson, and self-dual Pleba\'nski-Demia\'nski metrics. It is worth emphasizing the striking fact that the whole {\em curved} geometry is captured by a {\em flat} space structure, bypassing the standard deformation approach of the non-linear graviton construction and the associated, highly non-trivial, problem of finding a new family of holomorphic curves in twistor space~\cite{Penrose:1976js,Atiyah:1978wi,Gindikin:1986}.

An immediate corollary of our construction is that all self-dual black holes admit a single Kerr-Schild description, with explicit expressions for the corresponding null vectors and scalar fields encoded by the dual twistor quadric. While this feature was previously known for the Eguchi-Hanson and self-dual Taub-NUT cases, the fact that it also holds for self-dual Pleba\'nski-Demia\'nski is not only novel but particularly noteworthy, given the complexity of the metric \eqref{PDmetric}. Moreover, in all cases this feature emerges as a consequence of a unified framework that makes its geometric origin clear.

\medskip

The fact that the framework developed in this paper is fully adapted to a situation in which, for a fixed orientation, only one complex structure exists, is crucial for its future applications beyond self-duality. Indeed, a key feature of non-self-dual, astrophysical black holes is that they have only one complex structure (for a fixed orientation)~\cite{Flaherty:1974,Flaherty:1976}, in contrast to self-dual backgrounds where there is a 2-sphere worth of them. This intriguing fact seems to account for the many remarkable properties of black hole perturbations, such as the special geometry of the Teukolsky system and the existence of $\alpha$-surfaces in the perturbed space-time~\cite{Araneda:2018ezs,Araneda:2019uwy,Araneda:2022xii}.
The latter feature is a landmark of twistor constructions, but, given that the background is non-self-dual, standard twistor theory does not apply to this setting. Instead, it is a {\em two-} (rather than three-)dimensional twistor space that governs the system~\cite{Bailey:1991a,Bailey:1991b}, and this structure is precisely a (dual) twistor quadric such as the ones considered in this paper.

It follows that the perturbation theory of non-self-dual black holes is intimately connected to twistor quadrics, albeit the exact way in which the standard, remarkably powerful twistor tools should be modified to apply to this situation has remained obscure. The construction in this paper then provides an ideal framework to precisely understand and develop connections between two different approaches to gravitational perturbation theory: one based on the standard twistor theory of deformations of self-dual spaces, and another based on the Teukolsky system associated to a twistor quadric. The former is connected to many exciting developments such as the existence of chiral algebras related to soft expansions in quantum field theory~\cite{Guevara:2021abz,Strominger:2021mtt,Adamo:2021lrv}, while the latter constitutes the basis for analysis of black hole stability and gravitational wave physics (cf., \cite{Dafermos:2008en,Dafermos:2016uzj,Andersson:2016hmv,Dafermos:2017yrz,Andersson:2019dwi}). 
A concrete reason as to why such connections between the two approaches must exist is that all gravitational perturbations of self-dual spaces can be generated by the standard twistor construction~\cite{Hitchin:1980hp}, and at the same, all such perturbations are also expected to come from solutions to the Teukolsky equations. In particular, one immediate and intriguing question is then to understand how the known infinite-dimensional symmetry algebras are manifested in the Teukolsky system. Such questions are sure to provide a fruitful avenue for future research. 

\acknowledgments
	We thank James Lucietti for several helpful conversations. The authors are supported by a Royal Society University Research Fellowship (TA), the Simons Collaboration on Celestial Holography CH-00001550-11 (TA \& SS), the ERC Consolidator/UKRI Frontier grant TwistorQFT EP/Z000157/1 (TA \& BA) and the STFC consolidated grant ST/X000494/1 (TA).

\bibliographystyle{JHEP}
\bibliography{SDQ}

\end{document}